\documentclass{egpubl}
\usepackage{egsr2025_DLOnlyTrack}

\WsPaper           

\CGFStandardLicense

\usepackage[T1]{fontenc}
\usepackage{dfadobe}

\usepackage{cite}  
\BibtexOrBiblatex
\electronicVersion
\PrintedOrElectronic
\ifpdf \usepackage[pdftex]{graphicx} \pdfcompresslevel=9
\else \usepackage[dvips]{graphicx} \fi

\usepackage{egweblnk}

\usepackage{booktabs} 
\usepackage{xspace}
\usepackage{soul}

\usepackage{color}
\usepackage{graphicx}
\usepackage{amsmath}
\usepackage{multirow}
\usepackage{array}
\usepackage[capitalize,nameinlink]{cleveref}

\usepackage[english]{babel}
\usepackage[utf8]{inputenc}
\usepackage{algorithm}
\usepackage{algpseudocode}
\usepackage{overpic}
\usepackage{afterpage}

\usepackage{orcidlink}

\title{Stochastic Ray Tracing of Transparent 3D Gaussians}

\author[Xin Sun, Iliyan Georgiev, Yun (Raymond) Fei, Milo\v{s} Ha\v{s}an]{
    \parbox{\textwidth}{%
        \centering%
        Xin Sun\orcidlink{0000-0002-8710-2645} \quad
        Iliyan Georgiev\orcidlink{0000-0002-9655-2138} \quad
        Yun (Raymond) Fei\orcidlink{0000-0001-8553-1377} \quad
        Milo\v{s} Ha\v{s}an\orcidlink{0000-0003-3808-6092}
    }
    \vspace{-1mm}\\
    Adobe
    \vspace{-4mm}
}

\begin{document}


\newcommand{\todo}{{\color{red} TODO}}

\definecolor{XSColor}{rgb}{0.2,0.4,0.8}
\newcommand{\xs}[1]{{\color{XSColor} [\textbf{XS}: #1]}}

\definecolor{IGColor}{rgb}{0.4,0.8,0.2}
\newcommand{\ig}[1]{{\color{IGColor} [\textbf{IG}: #1]}}

\definecolor{MHColor}{rgb}{0.8,0.2,0.4}
\newcommand{\mh}[1]{{\color{MHColor} [\textbf{MH}: #1]}}

\definecolor{RFColor}{rgb}{0.8,0.4,0.2}
\newcommand{\rf}[1]{{\color{RFColor} [\textbf{RF}: #1]}}

\newcommand{\bx}{\mathbf{x}}
\newcommand{\ba}{\mathbf{a}}
\newcommand{\bb}{\mathbf{b}}
\newcommand{\bc}{\mathbf{c}}
\newcommand{\bm}{\mathbf{m}}
\newcommand{\bp}{\mathbf{p}}
\newcommand{\bq}{\mathbf{q}}
\newcommand{\br}{\mathbf{r}}
\newcommand{\bS}{\mathbf{S}}
\newcommand{\bR}{\mathbf{R}}

\newcommand{\bmu}{\boldsymbol{\mu}}
\newcommand{\bSigma}{\boldsymbol{\Sigma}}

\newcommand{\diag}{\text{diag}}
\newcommand{\fractional}{\text{fractional}}

\newcommand{\bear}{capture_07_toys_furry_toys_stuffed_toys_02242023}

\newcommand{\ouri}{OursMean\xspace}
\newcommand{\ourc}{OursCenter\xspace}

\maketitle


\begin{abstract}
3D Gaussian splatting has been widely adopted as a 3D representation for novel-view synthesis, relighting, and 3D generation tasks.
It delivers realistic and detailed results through a collection of explicit 3D Gaussian primitives, each carrying opacity and view-dependent color.
However, efficient rendering of many transparent primitives remains a significant challenge.
Existing approaches either rasterize the Gaussians with approximate per-view sorting or rely on high-end RTX GPUs.
This paper proposes a stochastic ray-tracing method to render 3D clouds of transparent primitives.
Instead of processing all ray-Gaussian intersections in sequential order, each ray traverses the acceleration structure only once, randomly accepting and shading a single intersection (or $N$ intersections, using a simple extension).
This approach minimizes shading time and avoids primitive sorting along the ray, thereby minimizing register usage and maximizing parallelism even on low-end GPUs. The cost of rays through the Gaussian asset is comparable to that of standard mesh-intersection rays.
The shading is unbiased and has low variance, as our stochastic acceptance achieves importance sampling based on accumulated weight. The alignment with Monte Carlo philosophy simplifies implementation and integration into a conventional path-tracing framework.
\end{abstract}


\begin{figure}[t]
    \includegraphics[width=1.0\linewidth]{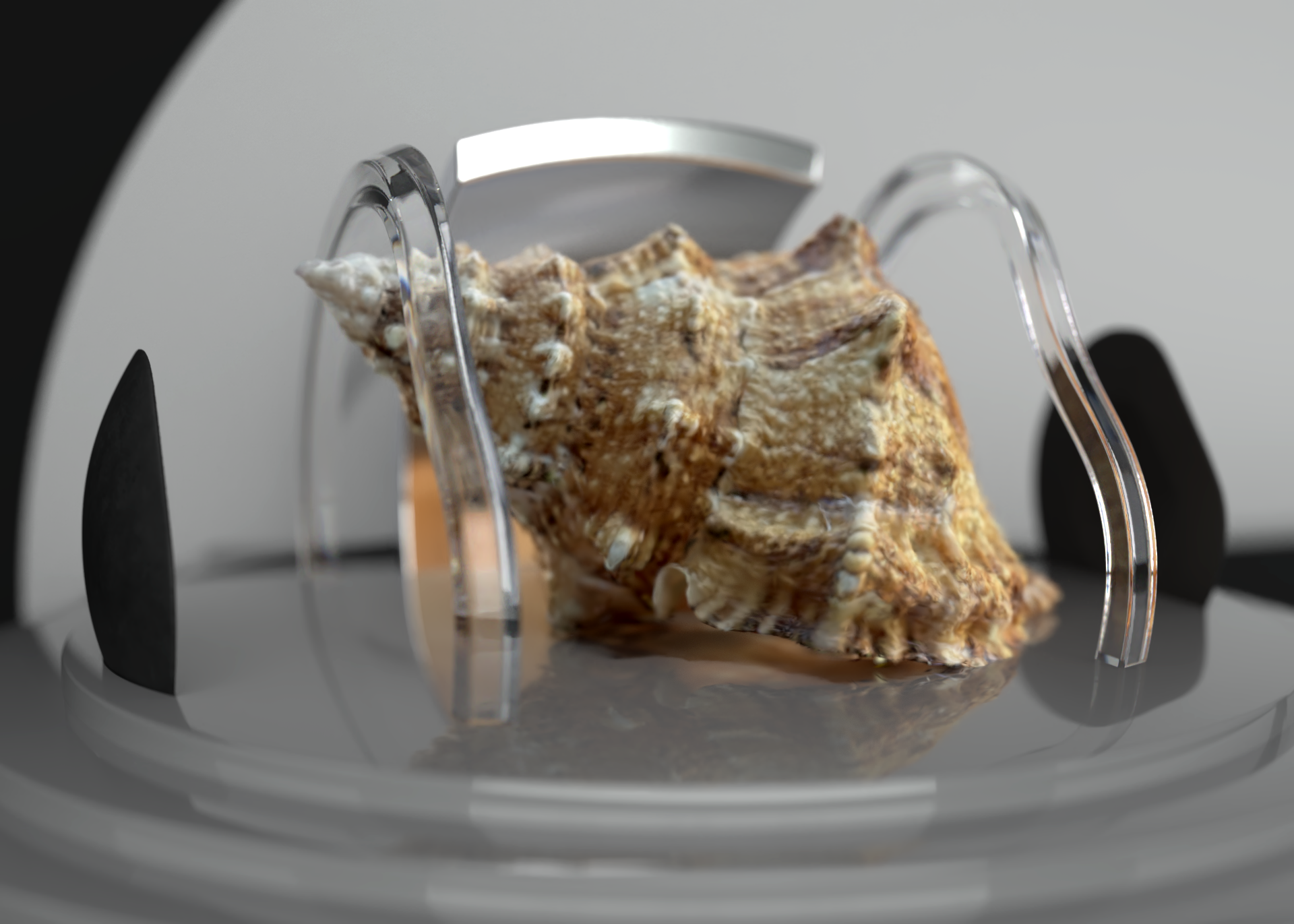}
    \vspace{-5mm}
    \caption{
        \textbf{Eternal Whisper of a Seashell.}
        A seashell reconstructed using 3D Gaussian splatting, in a scene made of traditional meshes and physically based materials. Shadows, glossy reflections on the base, refractions in curved glass, and depth-of-field effects are seamlessly added using a Monte Carlo path tracer that integrates our method. Please also see the supplementary videos.
    }
    \label{fig:seashell}
    \vspace{-3mm}
\end{figure}

\section{Introduction}
\label{sec:introduction}

Following the work of Kerbl et al.~\cite{kerbl20233dgs}, Gaussian splatting and its variations have become the de-facto standard 3D representation for novel-view synthesis, relighting, and 3D generation. This representation is based on a collection of explicit 3D Gaussians carrying opacities and view-dependent colors, and produces realistic and detailed reconstructions.

However, it is not obvious how to render a large set of scattered semi-transparent primitives accurately and efficiently.
Even for camera rays, many layers of partially visible primitives may contribute to the final shading, not to mention the cost of secondary effects (e.g., shadows and inter-reflections). This problem (and our solution) is not specific to 3D Gaussians and extends to any semi-transparent primitive that rays can intersect; for simplicity, we will assume 3D Gaussians in this paper.

Existing methods render Gaussians using either rasterization or ray tracing, and both approaches struggle with multi-layer transparency.
Rasterization requires sorting Gaussians per view or image bucket, which merely approximates the exact per-ray sorting and inevitably introduces errors that manifest as visual artifacts upon camera motion.
In addition, rasterization has inherent limitations in handling lighting effects such as shadows, reflections, and global illumination.
To address these issues, 3D Gaussian ray tracing \cite{3dgrt} has been proposed.
However, this method requires the sequential computation of all (relevant) Gaussian intersections along a ray, to ensure correct shading. Furthermore, it uses triangle meshes to bound Gaussian primitives, which enables the use of standard mesh acceleration structures for Gaussian ray tracing, but can be expensive on some hardware with limited resources since the number of triangles is a multiple of the number of Gaussians.

In this paper, we propose a stochastic ray-tracing method to render 3D clouds of transparent primitives. Instead of processing all ray-Gaussian intersections in sequential order, each ray traverses the acceleration structure only once, accepting and shading just a single intersection. As a simple extension, we also show how to shade $N$ intersections within a single traversal.

Unlike previous approaches, our method incorporates a stochastic decision inside the ray-traversal logic: each intersection is accepted probabilistically based on its opacity.
The fractional opacity of the intersection is treated as a probabilistic decision between a fully opaque and fully transparent event. We prove that this strategy yields an unbiased estimate of the final radiance.
It minimizes shading time as only the closest accepted intersection per ray needs shading.
It also avoids sorting the Gaussians along the ray and even storing extra data in its payload. In addition, once an intersection is accepted, the BVH nodes beyond the intersection can be skipped by clipping the ray segment, further lowering the traversal cost. In a GPU implementation, this method also saves on register usage, maximizing on-chip parallelism even on low-end GPUs.

While the stochastic nature of our method introduces noise, it converges rapidly over just a few iterations. In scenes containing both Gaussian assets and mesh geometry with traditional materials, we consistently observe that noise due to our method diminishes faster with increasing sample count than the variance due to more complex light paths. This supports the idea that a fast low-noise estimator is more beneficial than a slower noise-free one for ray-tracing Gaussians in such practical scenes. Thanks to its simplicity, our approach has been successfully integrated into a commercial Monte Carlo renderer, allowing seamless rendering of 3D Gaussians alongside conventional 3D assets (\cref{fig:seashell}).

\section{Related work}
\label{sec:relatedwork}

\paragraph*{Gaussian representations and applications.}

3D Gaussian splatting (3DGS) \cite{kerbl20233dgs} has become an established 3D representation for novel-view synthesis from multi-view images. It models objects and scenes as collections of thousands to millions of transparent anisotropic Gaussians with view-dependent color.

Many extensions to 3D Gaussian splatting have been proposed, e.g.\ to achieve better reflections \cite{deferred}, as well as relightable representations that store material properties per Gaussian \cite{r3dg,gsir}. Several methods for text-to-3D generation have also adopted 3D Gaussians as their output representation \cite{gslrm,trellis}. These applications are orthogonal to our work but call out the importance of fast and accurate rendering of the resulting clouds of transparent primitives.

\paragraph*{Addressing 3DGS limitations.}

The rasterization efficiency in 3DGS is key to its success, but it also comes with limitations: The 3D Gaussians are flattened into camera-facing splats (``billboards'') and their sorting order is also approximate. StopThePop \cite{stopthepop} addresses the first issue by using the mean of the 1D Gaussian along a virtual ray as the contribution point, and the second by a hierarchical sorting approach. Hahlbohm et al.~\cite{hahlbohm} introduced a hybrid approach with similar improvements. However, all rasterization approaches necessarily approximate the sorting order, or need to handle an unbounded number of primitives per pixel (see below).

2D Gaussian splatting \cite{2dgs} uses flat primitives with normal vectors, which can benefit the fitting of smooth surface structures and leads to a precise ray-Gaussian intersection definition.
Instead of splatting, Condor et al.~\cite{condor2024} treat mixtures of Gaussian  (or other, e.g.\ Epanechnikov) 3D kernels more rigorously as defining a volumetric density field that can be rendered using physically based volume-scattering approaches.
Exact volumetric ellipsoid rendering \cite{ever} uses 3D ellipsoids as another approach to turn a collection of transparent primitives into a rigorously defined volumetric field. These methods address the challenge of precisely defining the contribution of a transparent 3D primitive to a ray (pixel), but do not fundamentally increase the efficiency of handling many such primitives per ray.

\paragraph*{Order-independent transparency (OIT).}

OIT is the long-stan- ding problem of rasterizing unbounded numbers of partially transparent primitives with correctly ordered blending. The A-buffer \cite{abuffer} provides a correct solution but requires sorting unbounded arrays---a poor fit for modern GPU rasterization. Stochastic transparency \cite{stochtransp} addresses the issue using a Monte Carlo estimator at the cost of introducing some noise; we take a similar approach in the ray-tracing context. Multi-layer alpha tracing \cite{mlat} is a more recent method combining rasterization and ray tracing.

Two concurrent and independent works \cite{kheradmand2025stochasticsplats,hu2025real} propose methods closely related to ours. They share similar core ideas, applied in the context of efficient and accurate rasterization. Our method focuses on ray tracing,  demonstrating the effectiveness of the stochastic approach in scenarios involving other non-Gaussian assets, secondary reflections and soft shadows (see also supplementary videos).

To our knowledge, no prior work has explored the application of these ideas within the ray-tracing context, where it is typically assumed that sorting an arbitrary number of primitives is straightforward. While this assumption holds in principle, we demonstrate that relaxing strict sorting and instead adopting a Monte Carlo strategy---akin to stochastic transparency---can yield substantial efficiency improvements without compromising visual fidelity.

\paragraph*{Ray-tracing transparent primitives.}

R3DG~\cite{r3dg} proposed an inverse rendering method for relightable Gaussian reconstruction that includes a ray-tracing solution for visibility (transmittance) computation. A single BVH traversal finds all Gaussians along the ray and their transparencies are multiplied to compute the ray transmittance. This approach is related to ours but works only for transmittance where the intersection order does not matter. Our method is also based on a single BVH traversal but can compute unbiased radiance estimates, where order matters.

A concurrent work~\cite{wu20243dgut} unifies the representation for 3D Gaussian rasterization and ray tracing, rasterizing primary rays and ray-tracing secondary effects, significantly improving rendering performance. Our method fits into this formulation as it is not limited to tracing the splats used in the original 3DGS.

3D Gaussian ray tracing \cite{3dgrt} is the closest related work to ours. That approach bounds each Gaussian with a stretched icosahedron mesh and uses standard triangle-based ray-tracing acceleration structures to find the first $K$ primitives along the ray, repeating the tracing if more primitives are needed. Triangle ray tracing is well optimized on recent RTX GPUs, but this approach is not suitable for lower-end GPUs and CPUs.

\section{Stochastic ray tracing of transparency}
\label{sec:stochastic}

In this section, we explain our method in three steps. First, we define how a single primitive is handled along a ray. Second, we discuss how to quickly find and exactly handle all primitive intersections along a ray, assuming storage and sorting of the full array can be afforded. Finally, we present our Monte Carlo approach that avoids the overhead of storing and sorting the intersections.

\subsection{Handling a single Gaussian along a ray}

Our approach can handle any transparent primitives that can be (approximately) bounded and whose depth along a ray can be computed. For simplicity, we will assume a scene comprises a collection of 3D Gaussian primitives, each given as
\begin{equation}
    \label{eqn:gaussian}
    G(\bx) = e^{-\frac 1 2 \left(\bx - \bmu\right)^T\bSigma^{-1}\left(\bx - \bmu\right)},
    \quad
    \text{with}
    \quad
    \bSigma = \bR^{T}\bS^2\bR,
\end{equation}
where $\bmu$ is its mean and $\bSigma$ is its variance determined by a diagonal scaling matrix $\bS$ and a rotation matrix $\bR$.

While a 3D Gaussian is theoretically unbounded, we can compute an approximate axis-aligned bounding box (AABB) by bounding the ellipsoidal volume
\begin{align}
    \label{eqn:cuboid}
    \left\| \bS^{-1} \bR \left( \bx - \bmu \right) \right\|_2 \leq s,
\end{align}
where $s$ represents the standard deviation beyond which the Gaussian is considered negligible; in all our experiments we use $s = 2\sqrt{2} \approx 2.8$. For simplicity, we compute the bounding box of an unrotated Gaussian, rotate, and expand the box; a tighter bound can be found with more computation.

The intersection of a 3D Gaussian with a straight line (ray) is a 1D Gaussian along the line.
A natural way to define a intersection depth (i.e., shading position) is to take the mean of this 1D Gaussian \cite{3dgrt,stopthepop}; other definitions can also be used in our framework.
If the shading position lies behind the ray's origin, or outside the aforementioned bounding ellipsoid, the intersection is culled. The remaining intersected 3D Gaussians contribute to the final shading along the ray.

\subsection{Single BVH traversal with exact radiance computation}

As long as the AABBs of all 3D Gaussians are well defined, a spatial structure, such as a bounding volume hierarchy (BVH), can be efficiently constructed over them. This typically is done within frameworks such as Embree \cite{embree} and Optix \cite{optix}. A single ray traced through the scene will intersect multiple AABBs which can be found in a single BVH traversal; however, the resulting Gaussian intersections (if valid) will not be sorted in depth order. Instead, they will be in ``BVH order'' which roughly approximates depth order but could differ significantly in some cases, especially when large Gaussians are present.

A complex scene can contain millions of primitives, with a single ray potentially intersecting thousands of them. An exact solution would maintain a dynamic list of all intersections and sort the list before computing the radiance estimate (given below). This may be sufficient for some applications, but it also poses challenges for GPU implementation.

Given a ray-Gaussian intersection at depth $t$, we can evaluate its opacity $\alpha \in [0, 1]$ and shading color $c$.
The shading color may be dependent on the view or on other scene properties (e.g., material or lighting).
Given a ray intersecting $M$ 3D Gaussians, the exact shading color $L$ along the ray is the accumulated contribution from all intersections, sorted from closest to farthest:
\begin{align}
    \label{eqn:accumulated}
    L = \sum_{i=1}^M T_i \alpha_i c_i,
    \quad
    \text{with}
    \quad
    T_i = \prod_{j=1}^{i-1} (1 - \alpha_j),
\end{align}
where $T_i$ is the transmittance from all prior intersections along the ray.
$L$ can be interpreted as the foreground color, and the overall opacity along the ray is $1 - T_{M+1}$.
This result can be further composited with any background color to get the final rendering color.

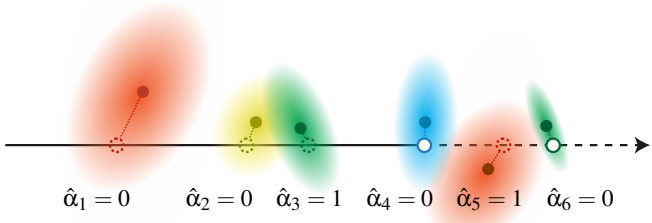
\begin{figure}[t]
    \begin{overpic}{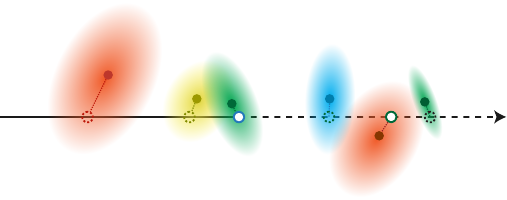}
        \put( 9,7){$\hat{\alpha}_1 = 0$}
        \put(28,7){$\hat{\alpha}_2 = 0$}
        \put(42,7){$\hat{\alpha}_3 = 1$}
        \put(56,7){$\hat{\alpha}_4 = 0$}
        \put(70,7){$\hat{\alpha}_5 = 1$}
        \put(84,7){$\hat{\alpha}_6 = 0$}
    \end{overpic}
    \vspace{-7mm}
    \caption{
        \textbf{Ray tracing stochastic binary opacities.}
        This example illustrates a ray intersecting $6$ 3D Gaussians.
        Each intersection has an opacity $\alpha_i$ at the mean of a 1D Gaussian.
        A random number $\xi_i \in [0, 1]$ determines the binary opacity $\hat{\alpha}_i$, where $\hat{\alpha}_i = 1$ if $\xi_i < \alpha_i$ and $\hat{\alpha}_i = 0$ otherwise.
        In this case, only $\hat{\alpha}_3$ and $\hat{\alpha}_5$ are opaque ($=1$); the rest are transparent ($=0$).
        Final shading uses the closest opaque intersection, $\hat{\alpha}_3$.
        If $\hat{\alpha}_5 = 1$ is accepted, farther intersections (e.g., $\alpha_6$) can be skipped.
        Evaluations need not follow distance order.
        For instance, if $\hat{\alpha}_5 = 1$ is stored in the ray's payload and $\hat{\alpha}_3 = 1$ is later accepted, the payload is simply updated with $\hat{\alpha}_3 = 1$.
        Different runs may yield different accepted intersections.
        Such runs can be performed simultaneously during a single traversal (\cref{sec:stochastic:multisample}).
    }
    \label{fig:illustration}
\end{figure}

\subsection{Stochastic binary opacities}
\label{sec:stochastic:opacity}

We are now ready to introduce our Monte Carlo approach which avoids the need to sort or store the intersections.
We introduce a Russian Roulette process to define binary opacities $\hat{\alpha}_i \in \{0, 1\}$:
\begin{align}
    \label{eqn:rr}
    \hat{\alpha}_i =
    \begin{cases}
        1, & \text{with probability $\alpha_i$,} \\
        0, & \text{with probability $1 - \alpha_i$}.
    \end{cases}
\end{align}
We have thus constructed $M$ binary random variables $\hat{\alpha}_i$ with expectations matching the opacities of the primitives along the ray: $E[\hat{\alpha}_i] = {\alpha}_i$. Since the shading $L$ depends linearly on each of the opacities $\alpha_i$ in isolation, we can replace every $\alpha_i$ by $\hat{\alpha}_i$ in \cref{eqn:accumulated} to obtain an unbiased estimator $\hat{L}$ for $L$:
\begin{align}
    \label{eqn:estimate}
    \hat{L} = \sum_{i=1} \hat{T}_i \hat{\alpha}_i c_i,
    \quad
    \text{where}
    \quad
    \hat{T}_i = \prod_{j=1}^{i-1} (1 - \hat{\alpha}_j).
\end{align}
The unbiasedness of $\hat{L}$ depends on the mutual independence of the random variables $\hat{\alpha}_i$. Specifically, if $\hat{\alpha}_i$ and $\hat{\alpha}_j$ are mutually independent, the expectation of their product is equal to the product of their individual expectations: $E[\hat{\alpha}_i \cdot \hat{\alpha}_j] = E[\hat{\alpha}_i] \cdot E[\hat{\alpha}_j] = \alpha_i \cdot \alpha_j$.

With this process, if an intersection is accepted as having an opacity of one, all subsequent intersections (i.e., ones with greater depths $t_j$) along the ray can be ignored.
As a result, the estimator $\hat{L}$ reduces to the contribution from the closest accepted intersection, denoted by index $i$ (see \cref{fig:illustration}). Formally:
\begin{equation}
    \label{eqn:cloest}
    \hat{L} = c_i,
    \quad
    \text{where}
    \quad
    \begin{cases}
        \hat{\alpha}_i = 1, \\
        \hat{\alpha}_j = 0, & \forall j \mbox{ such that } t_j < t_i.
    \end{cases}
\end{equation}

\subsection{Ray intersection algorithm}
\label{sec:stochastic:kernel}

Our method requires a single ray-BVH traversal operation that searches for the closest accepted intersection. For each processed primitive bounding box, the intersection routine in \cref{alg:intersect} is invoked. The intersection is accepted if it (1)~passes the Russian Roulette test, (2)~lies within the valid ray range, and (3)~is not negligible (i.e., is inside the bounding ellipsoid).
Upon acceptance, the callback reports the hit to update the ray's far range to exclude farther intersections. The reported hit, containing the primitive's index, will then be shaded in a ray-hit program.

With this approach, the ray is traced only once.
Most intersections are skipped, as only those closer than the latest accepted intersection are processed further.
Furthermore, each ray is shaded at most once, at the closest accepted intersection returned by the intersection routine in \cref{alg:intersect}.

\subsection{Efficient multi-sample estimation}
\label{sec:stochastic:multisample}

A single evaluation of the estimator $\hat{L}$ \eqref{eqn:cloest} per pixel can yield a noisy image.
To reduce this noise, the standard approach is to average the contributions of $N$ evaluations:
\begin{align}
    \label{eqn:multisample}
    \hat{L} = \frac{1}{N} \sum_{k=1}^N \hat{L}_k.
\end{align}
This can be achieved by tracing a ray independently $N$ times. Our method allows for performing such $N$ instantiations (for one ray) within a \emph{single} BVH traversal. For each instantiation, we track the ID and hit distance of the closest accepted primitive. And for each tested primitive, we instantiate the stochastic opacity $\hat{\alpha}_i$ $N$ times. At the end of the traversal, we have $N$ independent (but not necessarily unique) intersections, each giving rise to an estimator $\hat{L}_k$.
This implementation can be very efficient, provided that the increased memory usage does not significantly degrade on-chip parallelism.

\begin{algorithm}[t]
    \small
    \caption{
        Ray-primitive intersection routine. Inputs: 3D Gaussian $g$ with bounding box intersected by the ray $r$.
    }
    \label{alg:intersect}
    \begin{algorithmic}[1]
        \Procedure{IntersectPrimitive}{$g$, $r$}
            \State $g_1 \gets \Call{GetGauss1D}{g, r}$ \Comment{Compute 1D Gaussian along ray}
            \State $t \gets g_1.\mu$ \Comment{Retrieve the 1D Gaussian mean}

            \If{$t \leq r.t_\mathrm{min}$ \textbf{or} $t \geq r.t_\mathrm{max}$}
                \State \Return \Comment{Intersection is outside the valid ray range}
            \EndIf

            \If{$\Call{IsNegligible}{g, g_1}$}
                \State \Return \Comment{Mean of $g_1$ is outside the AABB of $g$; \cref{eqn:cuboid}}
            \EndIf

            \State $\boldsymbol{p} \gets r.\boldsymbol{o} + t \ r.\boldsymbol{d}$ \Comment{Calculate intersection position}

            \State $\xi \gets \Call{RNG}{0, 1, \boldsymbol{p}}$ \Comment{Position-dependent pseudo-random number}

            \If{$\xi < g_1.\alpha$}
                \State \Return \Comment{Intersection rejected by Russian Roulette}
            \EndIf

            \State $\Call{ReportIntersection}{g, t}$ \Comment{Clip ray: $r.t_\mathrm{max} \gets t$}
        \EndProcedure
    \end{algorithmic}
\end{algorithm}

\subsection{Discussion}

A notable advantage of our stochastic approach is its computational simplicity on GPUs.
Unlike previous approaches, our single-sample variant does not require maintaining a dynamic buffer of intersections in registers or global video memory, nor need repeated ray generations and traversals.
Buffers in global memory suffer from high access latency, while per-thread fixed buffers consume additional registers, reducing on-chip parallelism.
This issue is particularly acute on low-end GPUs with more constrained resources.

In our method, the ray payload remains minimal as it only stores the nearest ``opaque'' intersection.
This is also advantageous in graphics APIs such as Vulkan and DXR, which impose strict limitations on direct access to the ray payload (a register-based, per-path data structure) in an intersection shader, for storing a long list of samples. Instead, these APIs often require round-trips between any-hit and intersection shaders, increasing instruction count and implementation complexity.

When the camera stops moving, a few additional iterations may still be needed for the image to fully converge, but this delay becomes negligible when 3D Gaussians are rendered alongside other types of 3D assets in a Monte Carlo path tracing framework.
3D Gaussians contribute to global illumination effects such as shadows and reflections, with stochastic opacity being only one source of Monte Carlo noise among various other sampling processes. In scenes featuring complex materials, we frequently observe that the convergence on stochastic opacity often occurs earlier than for other effects.
That said, in such scenarios, quickly obtaining a result from a ray is more beneficial than perfectly shading the Gaussians in every iteration, as the latter's higher computational cost can slow down overall scene convergence. Alternatively, improved convergence can be achieved through a global importance sampling strategy for each light path sample, rather than focusing exclusively on noise-free rendering of 3D Gaussians.

Finally, with the cost of storing some samples in the ray payload, the multi-sample variation provides an option to balance between interactivity and faster convergence.

\section{Implementation details}
\label{sec:implementation}

\subsection{Matching rasterizer's depth estimate}
\label{sec:implementation:depth}

Using the mean of the 1D Gaussian can lead to results that are inconsistent with those produced by a rasterizer.
This discrepancy arises because the original 3DGS calculates depth based on the projected center of 3D Gaussians onto the camera direction, and interpolates inverse depth after screen projection.

It would be best to train the Gaussians with mean-based depth \cite{stopthepop,3dgrt}.
However, for compatibility with existing 3D assets reconstructed using publicly available rasterizer-based tools (e.g., PolyCam, Scaniverse), we can adapt the ray tracer to approximately align with center convention.
Specifically, the distance $t$ in \cref{alg:intersect} is computed by projecting the Gaussian center onto the camera direction. This still does not exactly match the rasterizer, because a Gaussian does not remain a Gaussian under an affine transformation. As our results show, the remaining error is minor.

\subsection{Stateless GPU pseudo-random number generation}
\label{sec:implementation:rng}

Some commonly used GPU (pseudo-)random number generators (e.g., a Sobol) usually have a state initialized with a seed, and need to update the state for the next generated number. Nevertheless, some modern graphics APIs for GPU raytracing, like Vulkan or DXR, do not allow writing to a ray payload or buffers in an intersection shader to update this state. Instead, with these APIs, one has to update in a special any-hit shader, which in turn requires routing the ray-tracing data back and forth between different shaders to determine the acceptance. Such a requirement introduces extra cost and forbids a unified ray-tracing framework across various platforms: for example, Metal instead does not allow an any-hit shader.

To mitigate these issues, we use a canonical stateless trigonometric hash function~\cite{rey1998random} to generate pseudo-random numbers in the intersection shader. Although other hash functions exist with higher sampling quality~\cite{jarzynski2020hash}, we use the trigonometric one as it is called frequently (for each potential intersection), so high efficiency is crucial. In addition, to make the hashed numbers frame-dependent, we hash them on the hit position (\cref{alg:intersect}, line 11) which has been perturbed with frame-number-dependent quasi-random numbers sampled from a stateful Sobol sequence~\cite{bradley2011parallelization} during the camera-ray generation. Thus, the numbers generated by hashing are also frame-number-dependent.

In particular, we use two hash functions, one scalar and one 2D:
\begin{align}
    \label{eqn:hash}
    r_1(q) = \fractional(b_1\text{sin}(a_1q)),\\
    \br_2(\bq) = \fractional(\bb_2\text{sin}(\ba_2^{\text{T}}\bq)),
\end{align}
where ``$\fractional()$'' takes the value's fractional part.
The heuristically chosen coefficients $a_1$, $\ba_2$, $b_1$, and $\bb_2$ are large enough so that the trigonometric function has sufficiently high frequency. For a 3D hit position $\bp$, we generate a random number (see \cref{alg:intersect})
\begin{align}
    \label{eqn:rng}
    \xi(\bp) = \br_2(\bp_{xy} + r_1(\bp_z)),
\end{align}
where $\bp_{xy}$ is the 2D vector containing the $x$- and $y$-coordinates of $\bp$ and $\bp_z$ is the $z$-coordinate of $\bp$.
One particular feature we want from the trigonometric hash function is effectively enlarging the perturbation to produce sampling with sufficient quality. Notice that the hit position $\bp$ can be treated as the hit position of the ray without perturbation (i.e., a function of the geometry and screen pixel location), plus the additive quasi-random perturbation as a function of both the stateful Sobol quasi-random number generator and the geometry. With sufficiently large $a_1$, $\ba_2$, $b_1$, and $\bb_2$, the sine function applied to $\bp$ will effectively enlarge the disturbance and eliminate the dependence on the regularity of the screen pixel position due to its cyclical nature.

We use $a_1=91.3458$, $\ba_2=[12.9898,78.233]^T$, $b_1=47453.5453$, $\bb_2=[43758.5453,43758.5453]^T$. The quasi-random number $\xi$ is sufficiently uniform for the ray tracer to converge.

\begin{table}[t]
    \centering
    \small
    \caption{
        \textbf{Offline rendering performance} (in seconds).
        $\#G$ means number of 3D Gaussians in millions.
        We test the performance of \ouri on a Windows 11 desktop, implemented with Vulkan on GPU and Embree on CPU.
    }
    \setlength\tabcolsep{1.65mm}%
    \begin{tabular}{c|c|cccccc}
        \toprule
        \multirow{2}{*}{\textbf{Asset}} & \multirow{2}{*}{$\mathbf{\#G}$}  & \multicolumn{2}{c}{\textbf{64\,spp}}      & \multicolumn{2}{c}{\textbf{256\,spp}} & \multicolumn{2}{c}{\textbf{1024\,spp}}    \\
        & & \textbf{GPU} & \textbf{CPU} & \textbf{GPU} & \textbf{CPU} & \textbf{GPU} & \textbf{CPU} \\
        \midrule
        \textit{drjohnson}  & 3.41          & 0.86 & 46.78    & 3.35 & 191.33   & 13.41 & 759.40  \\
        \textit{playroom}   & 2.55          & 0.60 & 31.68    & 2.43 & 127.39   &  9.32 & 511.90  \\
        \textit{room}       & 1.59          & 0.61 & 31.93    & 2.38 & 126.95   &  9.43 & 508.62  \\
        \midrule
        \textit{furniture}  & 0.11          & 0.27 & 7.07     & 1.60 & 28.37    & 5.49 & 116.72  \\
        \textit{cart}       & 0.18          & 0.27 & 12.13    & 1.60 & 48.64    & 6.30 & 195.45  \\
        \textit{girl}       & 0.15          & 0.40 & 12.40    & 1.60 & 49.60    & 7.01 & 197.49  \\
        \textit{racoon}     & 0.09          & 0.53 & 7.47     & 1.28 & 28.91    & 5.38 & 116.93  \\
        \midrule
        \textit{bear}       & 0.42          & 1.47 & 10.00    & 1.81 & 39.79    & 7.42 & 159.09  \\
        \textit{jacket}     & 0.31          & 0.53 & 12.00    & 2.03 & 47.25    & 7.72 & 189.66  \\
        \textit{shoe}       & 0.48          & 0.53 & 12.40    & 2.13 & 49.49    & 8.33 & 197.79  \\
        \textit{armor}      & 2.90          & 0.53 & 8.53     & 2.03 & 33.60    & 8.53 & 134.40  \\
        \midrule
        \textit{sphere}     & 2.24          & 0.80 & 39.73    & 3.20 & 162.03   & 12.80 & 656.46 \\
        \textit{sculpture}  & 7.63          & 0.93 & 40.40    & 4.16 & 157.12   & 16.66 & 639.80 \\
        \textit{bike}       & 5.85          & 1.07 & 45.20    & 4.27 & 182.40   & 18.08 & 739.86 \\
        \textit{motorcycle} & 6.85          & 1.33 & 40.27    & 3.84 & 163.84   & 15.75 & 657.68 \\
        \bottomrule
    \end{tabular}
    \label{tab:perf-offline-windows}
\end{table}

\section{Results}
\label{sec:results}

We implemented our method on different platforms and within multiple graphics APIs.
On Windows 11, we integrated the method in a Vulkan \cite{vulkan} GPU path-tracing framework. We also tested a version using Embree \cite{embree} for CPU path tracing.
We tested on a desktop with an AMD Ryzen 9 5950X 16-Core Processor \@ 3.40 GHz CPU, 128 GBytes RAM, and an Nvidia GeForce RTX 3090 with 24 GB Video RAM.
On MacOS, we used Metal \cite{metal2025} to implement GPU path tracing and Embree for CPU path tracing.
We tested on a Macbook Pro 16-inch M1 Max with 32 GBytes RAM, running MacOS 14.7.1.

We tested our method with two different definitions for the depth of a Gaussian intersection (see \cref{sec:implementation:depth}).
When defining depth as the 1D Gaussian mean, we denote our method as \ouri.
When the depth is evaluated with respect to the center of the intersected 3D Gaussian, which is closer to original 3DGS, we denote our method as \ourc.
Assets from 3DGS \cite{kerbl20233dgs} are rendered at $1200\times800$ resolution; other assets are rendered at $1280\times960$.

\paragraph*{Performance.}

We measured the performance of our method on Windows (CPU and Vulkan, shown in \cref{tab:perf-offline-windows}) and MacOS (CPU and Metal, \cref{tab:perf-offline-macos}), with sampler per pixel (spp) ranging from 64 to 1024. We also measured the interactive rendering performance in frames per second and compared to 3DGS rasterization in \cref{tab:perf-interactive}. Please see the captions of the respective tables for more details.

\begin{table}
    \centering
    \small
    \caption{\textbf{Offline rendering performance on MacOS (in seconds).}
        We test the performance of \ouri on an M1 Macbook Pro, implemented using Metal on GPU and Embree on CPU.
        The GPU implementation brings up to $5\times$ speedup over CPU.
        Nevertheless, the GPU performs worse in some large scenes (e.g., $sculpture$). This is due to the excessive memory access, cache thrashing, and fragmented memory access during the BVH traversal, which overwhelm the GPU's bandwidth and parallel architecture, causing low arithmetic intensity.
        For example, comparing $sculpture$ with $jacket$, whose file-size difference is more than $20\times$, we observed a $1.6\times$ last-level cache miss rate, $52\times$ more cache bytes read, and more time spent on memory address translation (23.9\% vs. 7.19\%).
        The Mac CPU handles large assets better with more advanced caching, pre-fetching, and flexibility for irregular workloads.
    }
    \setlength\tabcolsep{1.7mm}%
    \begin{tabular}{c|cccccc}
        \toprule
        \multirow{2}{*}{\textbf{Asset}} & \multicolumn{2}{c}{\textbf{64\,spp}} & \multicolumn{2}{c}{\textbf{256\,spp}} & \multicolumn{2}{c}{\textbf{1024\,spp}} \\
        & \textbf{GPU} & \textbf{CPU} & \textbf{GPU} & \textbf{CPU} & \textbf{GPU} & \textbf{CPU} \\
        \midrule
        \textit{drjohnson}  & 19.5 & 72.3   & 72.6 & 283.9    & 285.1 & 1130.2    \\
        \textit{playroom}   & 13.3 & 49.6   & 47.8 & 193.1    & 185.9 & 767.0     \\
        \textit{room}       & 13.0 & 51.2   & 46.7 & 199.4    & 181.4 & 792.0     \\
        \midrule
        \textit{furniture}  & 6.8 & 10.4    & 21.6 & 36.2     & 81.1 & 139.5      \\
        \textit{cart}       & 4.3 & 17.7    & 11.8 & 65.4     & 41.8 & 256.2      \\
        \textit{girl}       & 5.4 & 18.4    & 16.2 & 68.4     & 59.4 & 268.2      \\
        \textit{racoon}     & 2.8 & 9.2     & 5.6 & 31.2      & 17.0 & 119.4      \\
        \midrule
        \textit{bear}       & 4.0 & 14.3 & 13.4 & 54.3 & 51.8 & 225.9 \\
        \textit{jacket}     & 4.4 & 16.6 & 15.6 & 65.8 & 60.6 & 263.1 \\
        \textit{shoe}       & 5.0 & 18.3 & 17.7 & 73.5 & 69.0 & 288.1 \\
        \textit{armor}      & 4.8 & 11.7 & 15.7 & 47.1 & 60.7 & 181.0 \\
        \midrule
        \textit{sphere}     & 16.1 & 66.7   & 58.8 & 261.4    & 229.8 & 1040.2    \\
        \textit{sculpture}  & 7769.8 & 67.6 & 31073.8 & 265.0 & 124289.8 & 1054.6 \\
        \textit{bike}       & 84.5 & 77.0   & 332.6 & 302.4   & 1325.0 & 1204.2   \\
        \textit{motorcycle} & 67.5 & 67.9   & 264.7 & 266.3   & 1053.3 & 1059.7   \\
        \bottomrule
    \end{tabular}
    \label{tab:perf-offline-macos}
\end{table}

\paragraph*{Comparison to 3DGS rasterization.}

\begin{figure}[t]
    \centering
    \rotatebox{90}{\makebox[0.20\columnwidth][c]{\textit{drjohnson}}}
    \includegraphics[width=0.30\columnwidth]{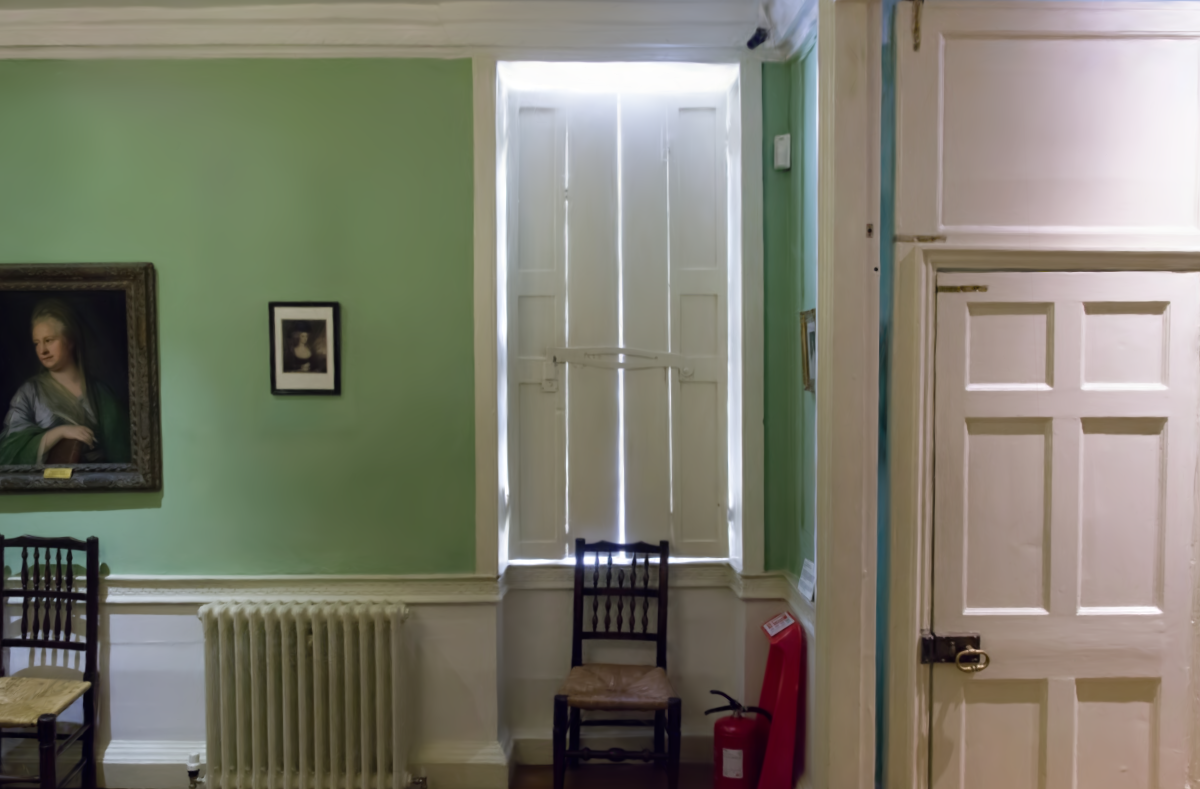}
    \includegraphics[width=0.30\columnwidth]{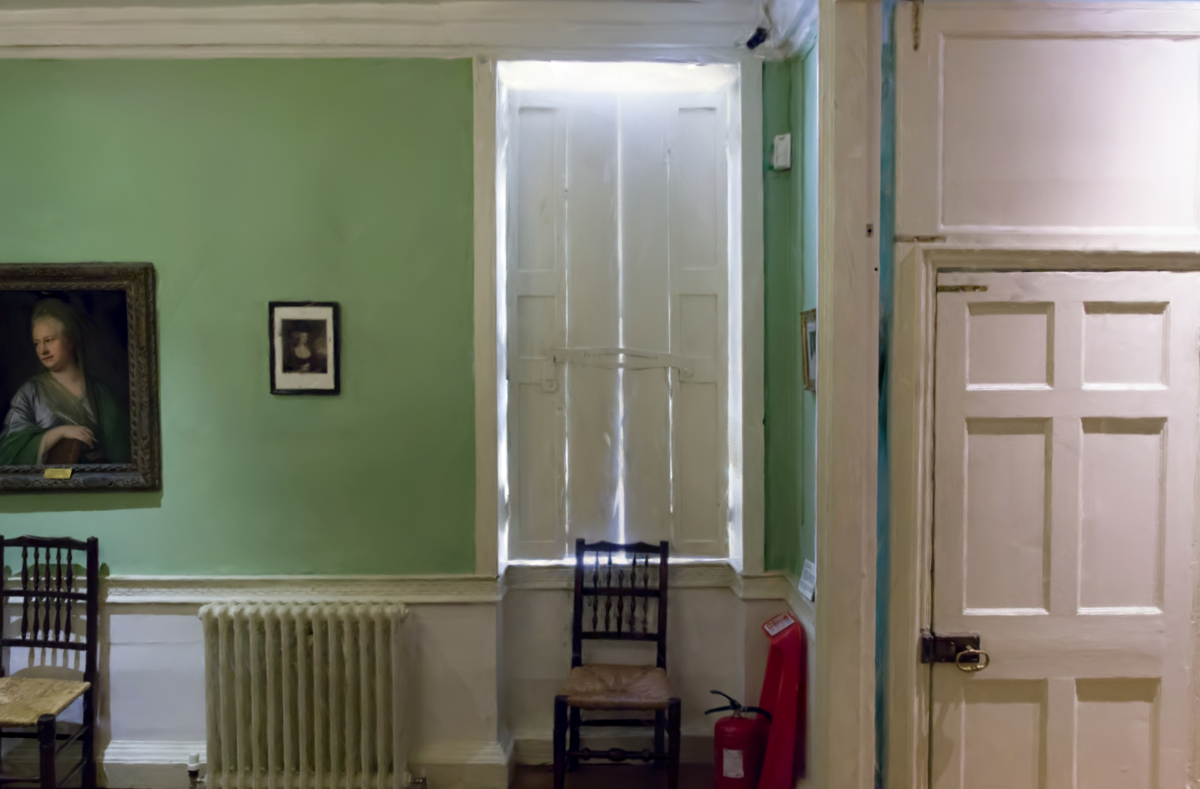}
    \includegraphics[width=0.30\columnwidth]{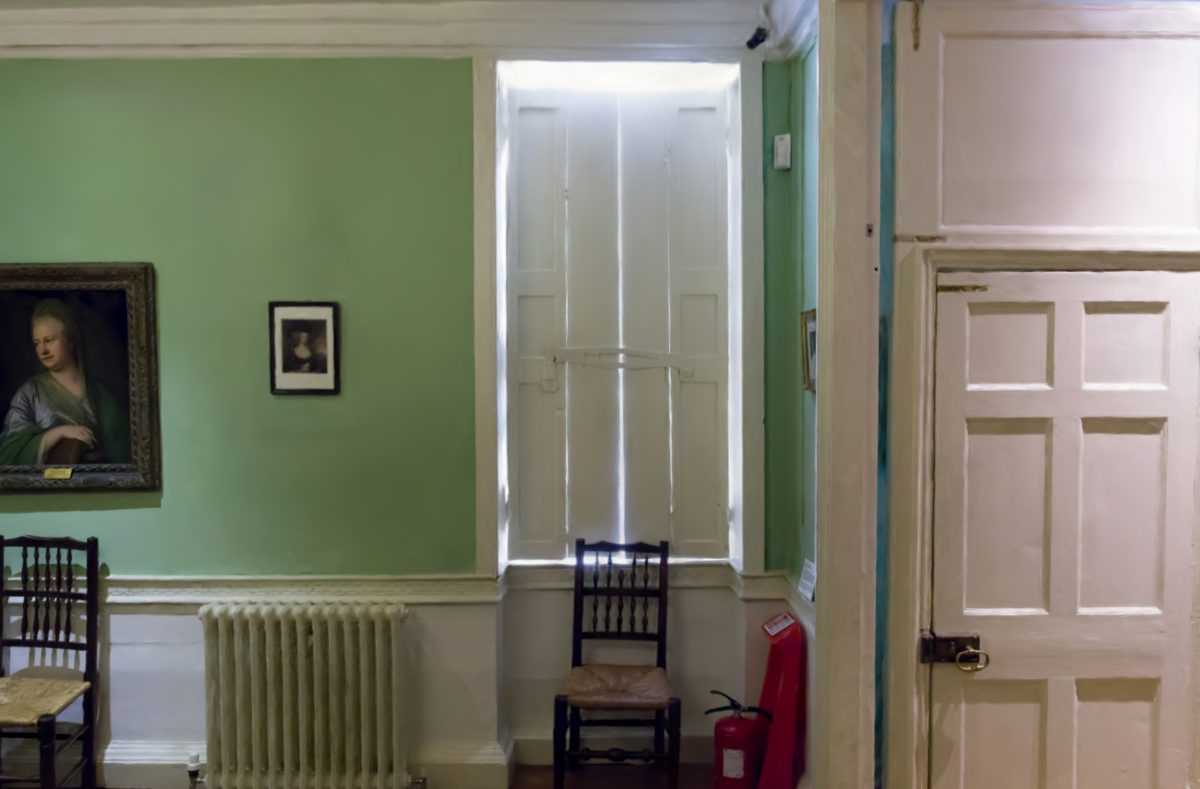}
    \\
    \rotatebox{90}{\makebox[0.20\columnwidth][c]{\textit{playroom}}}
    \includegraphics[width=0.30\columnwidth]{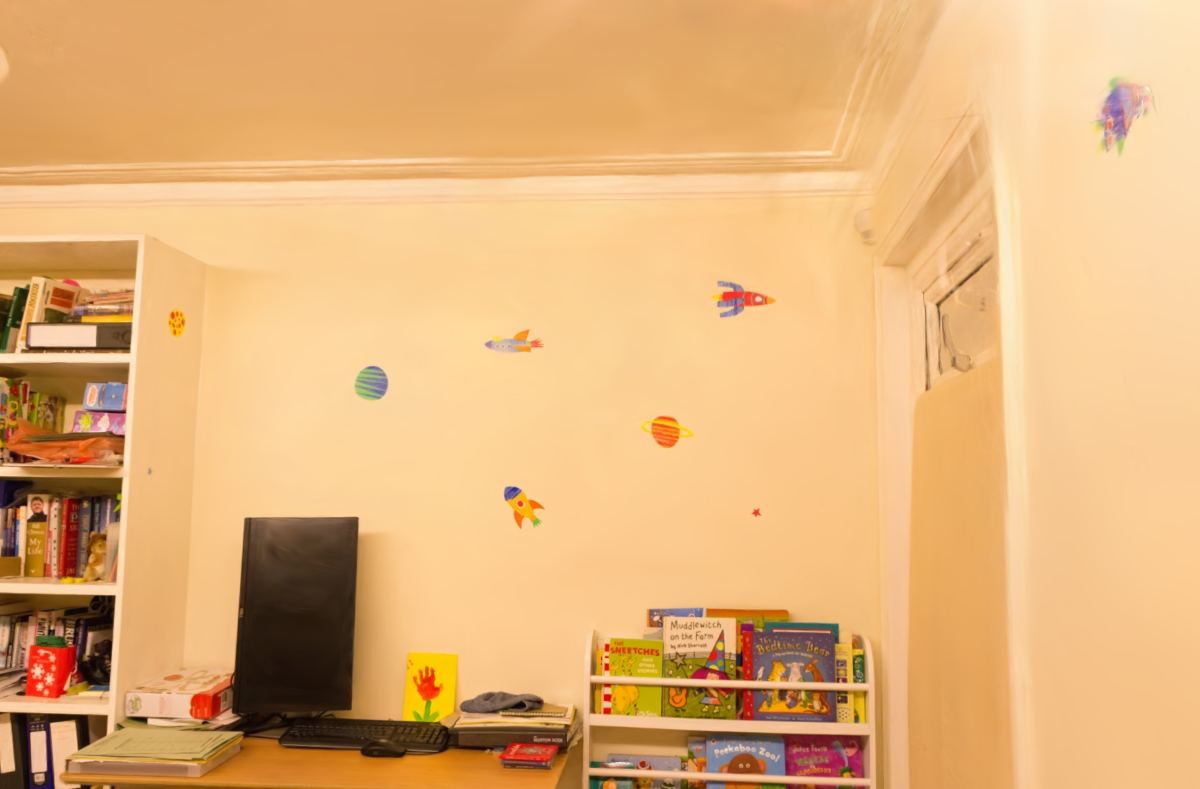}
    \includegraphics[width=0.30\columnwidth]{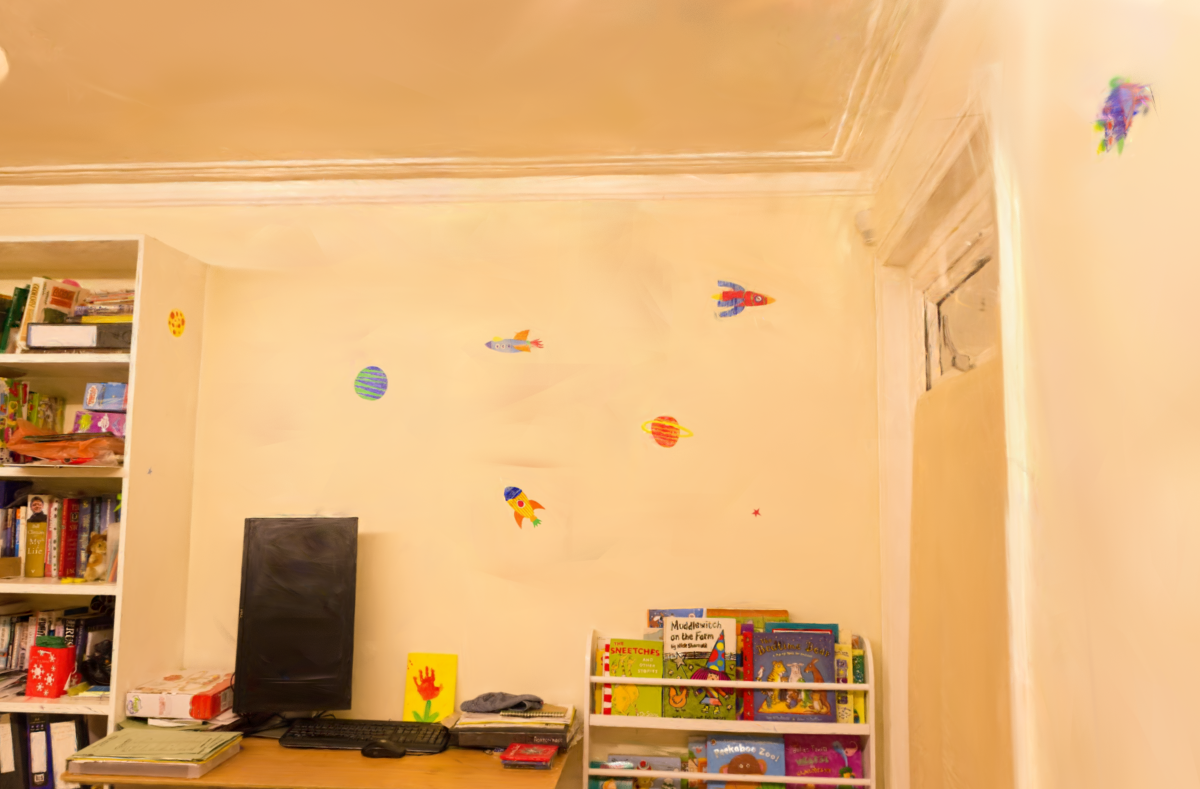}
    \includegraphics[width=0.30\columnwidth]{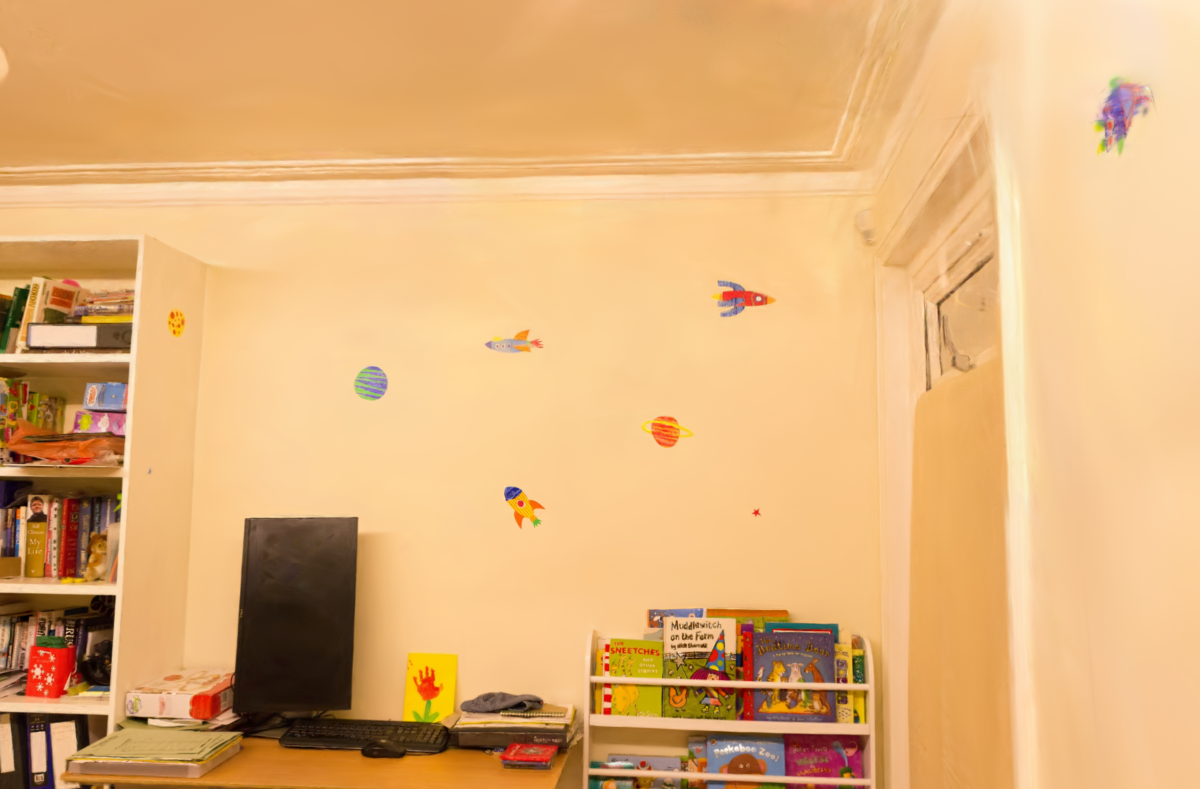}
    \\
    \rotatebox{90}{\makebox[0.20\columnwidth][c]{{\color{white}\textit{l}}\textit{room}\color{white}\textit{p}}} 
    \includegraphics[width=0.30\columnwidth]{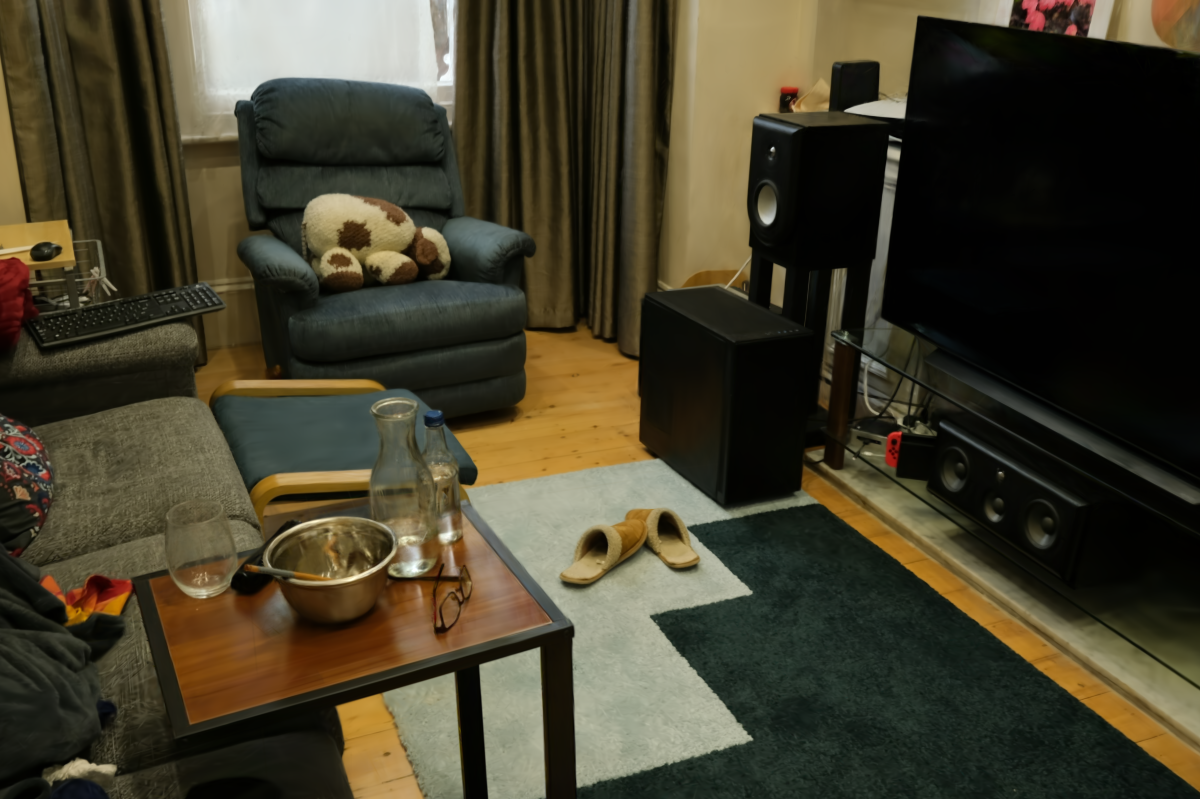}
    \includegraphics[width=0.30\columnwidth]{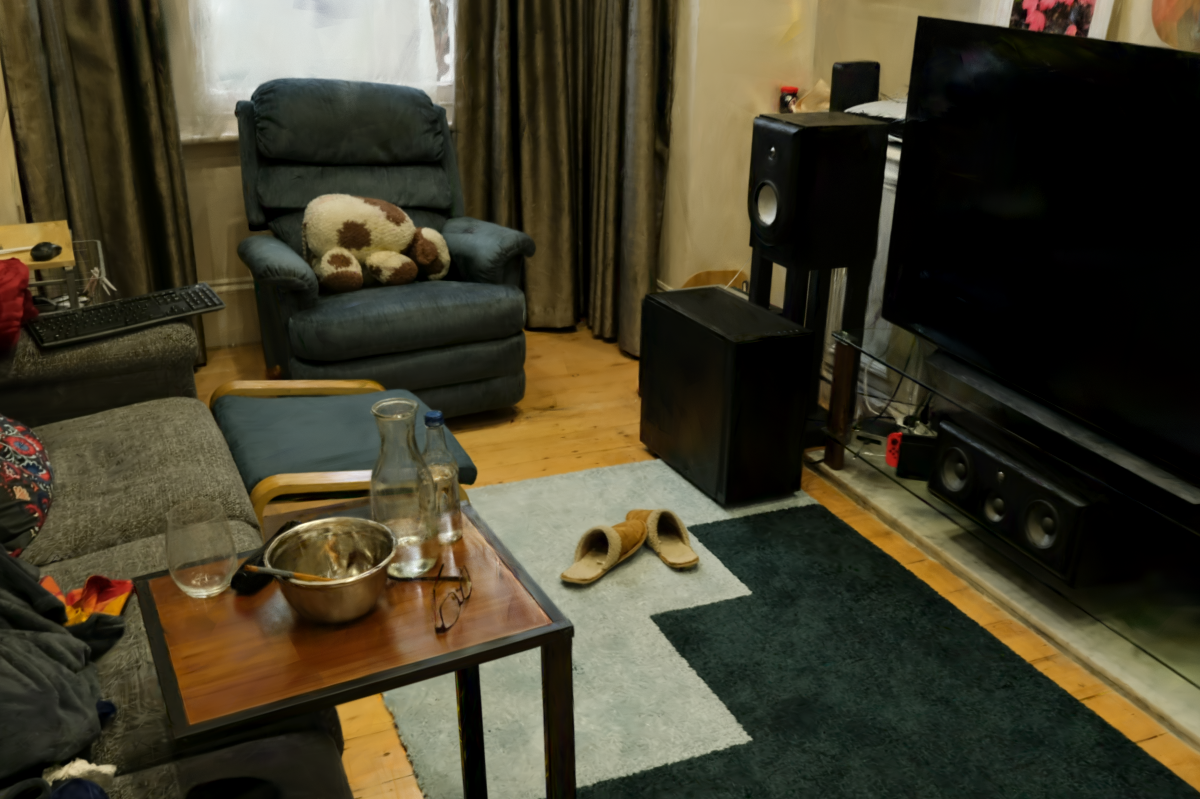}
    \includegraphics[width=0.30\columnwidth]{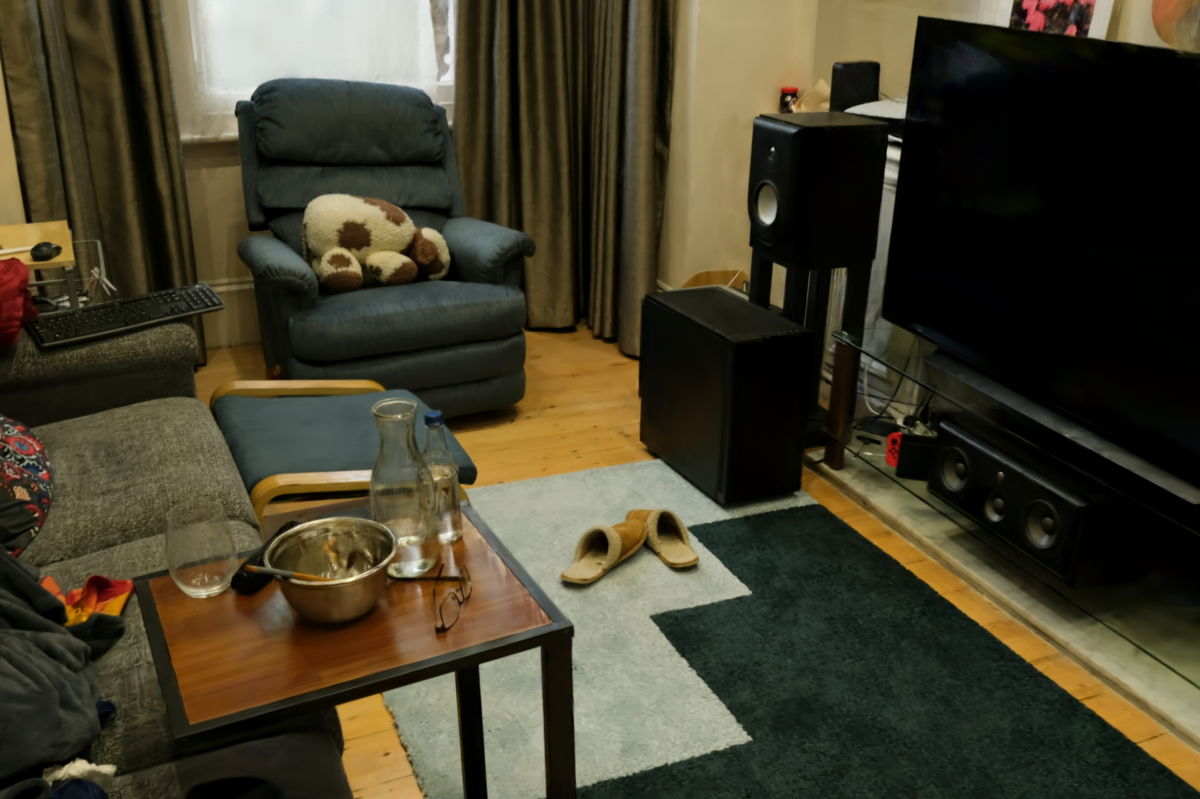}
    \\
    \rotatebox{90}{\makebox[0.01\columnwidth][c]{{\color{white}\textit{l}}\color{white}\textit{p}}} 
    \makebox[0.30\columnwidth][c]{\cite{kerbl20233dgs}}
    \makebox[0.30\columnwidth][c]{\textbf{\ouri}}
    \makebox[0.30\columnwidth][c]{\textbf{\ourc}}
    \caption{
        \textbf{Comparison with rasterization.}
        The three scene assets are from 3D Gaussian splatting \cite{kerbl20233dgs}, and are rendered with their open-sourced implementation in the leftmost column. The right two columns are both rendered with our stochatic ray tracing method. Because the assets are reconstructed using rasterization, we see some quality degradation in the second column because we evaluate the depth at the position where rays intersect 3D Gaussians. As introduced in \cref{sec:implementation:depth}, we adapt the depth according to the projected center of the intersected Gaussians in the rightmost column, producing images closely matching rasterization.
    }
    \label{fig:comparison-inria}
    \vspace{-2mm}
\end{figure}

In \cref{fig:comparison-inria}, we compare three assets from the 3DGS \cite{kerbl20233dgs}, showing their open-source implementation in the leftmost column. The right two columns are both rendered with our ray tracer. When we evaluate the depth at the 1D means, we see some quality degradation in the second column; this is because the assets are reconstructed using the center-based depth in the rasterizer. When we define depth according to the center of the intersected Gaussians in the rightmost column, our images match rasterization closer, as expected.

\paragraph*{Comparison to 3DGRT.}

3DGRT \cite{3dgrt} takes a very different approach tailored to RTX GPUs, making comparison non-trivial. However, even on RTX our method still has better single-sample performance. We evaluated the room scene in \cref{fig:comparison-inria} using the official 3DGRT implementation \cite{3dgrt} on a dual-boot Windows/Ubuntu22 system with an RTX 3090 GPU. The scene was trained using 3DGRT, resulting in an asset with $1.28$ million Gaussians, and rendered at $1557 \times 1038$ resolution. 3DGS rasterization renders at 3.9ms/frame (Windows \& Ubuntu). Ours renders at 12.2ms/frame (Windows); the performance is estimated by averaging over 64 spp. 3DGRT renders at 34.3ms/frame (Ubuntu).

\paragraph*{Assets from different sources.}

\begin{figure*}[t]
    \centering
    \includegraphics[width=0.500\columnwidth]{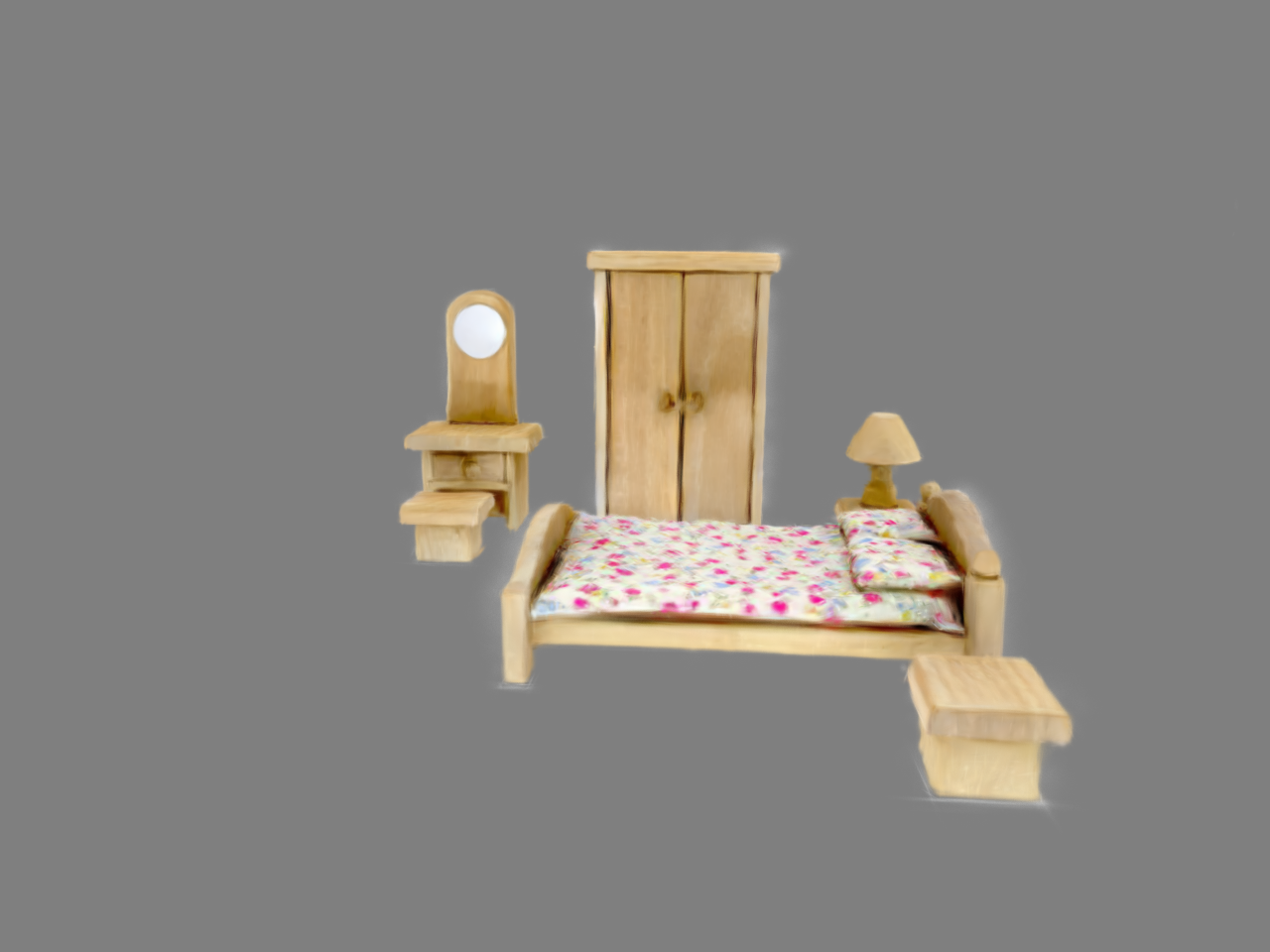}
    \includegraphics[width=0.500\columnwidth]{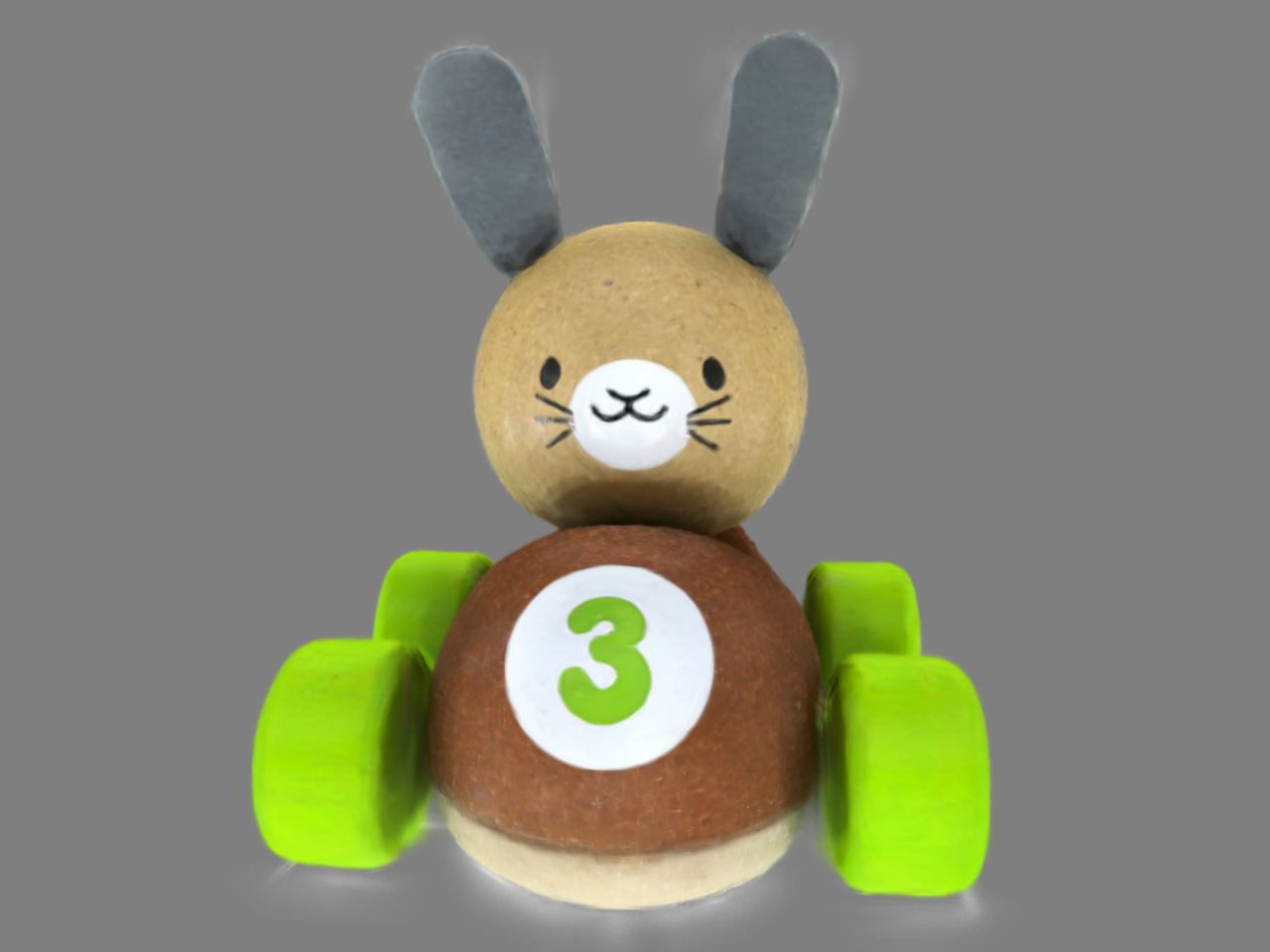}
    \includegraphics[width=0.500\columnwidth]{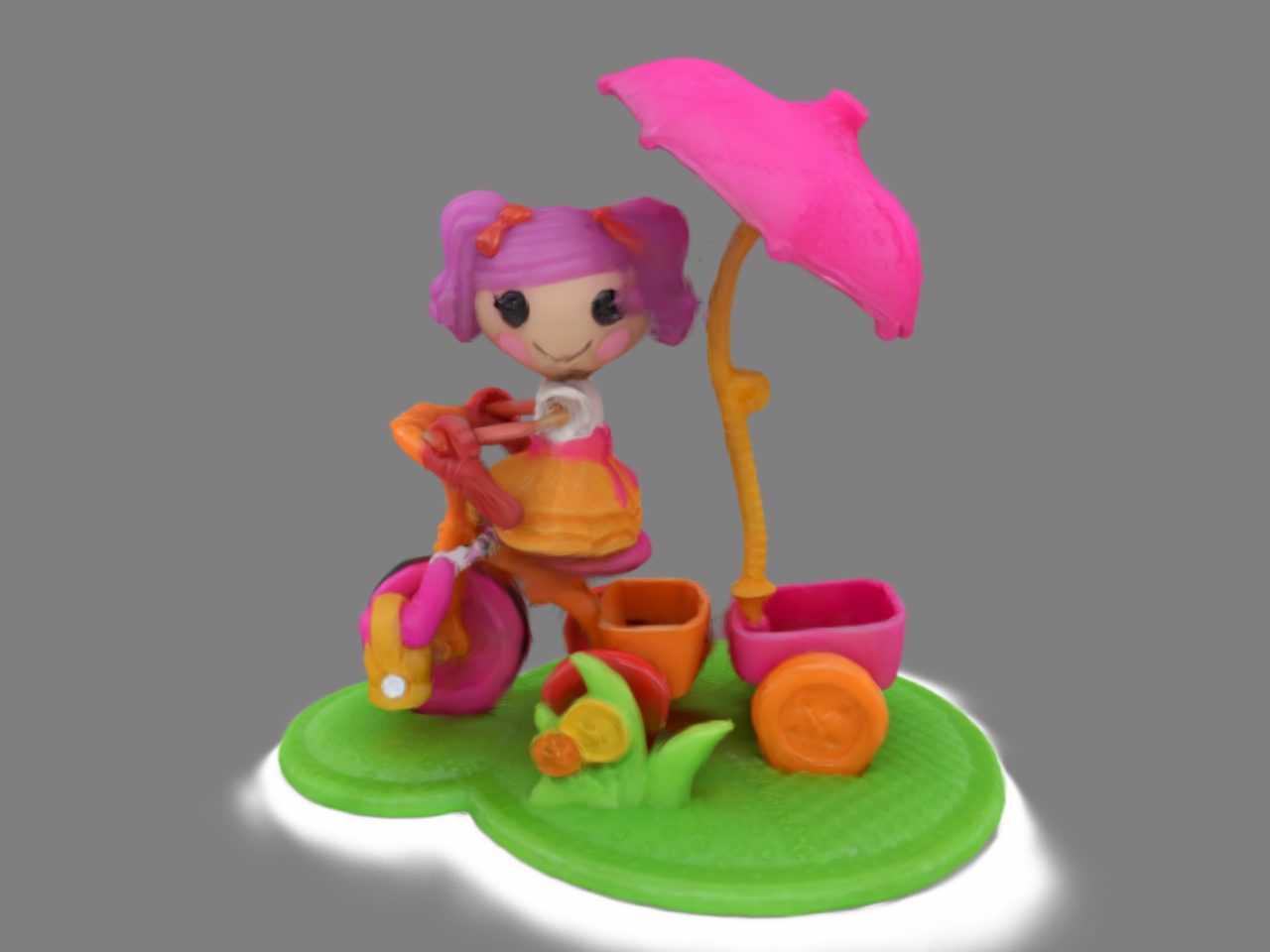}
    \includegraphics[width=0.500\columnwidth]{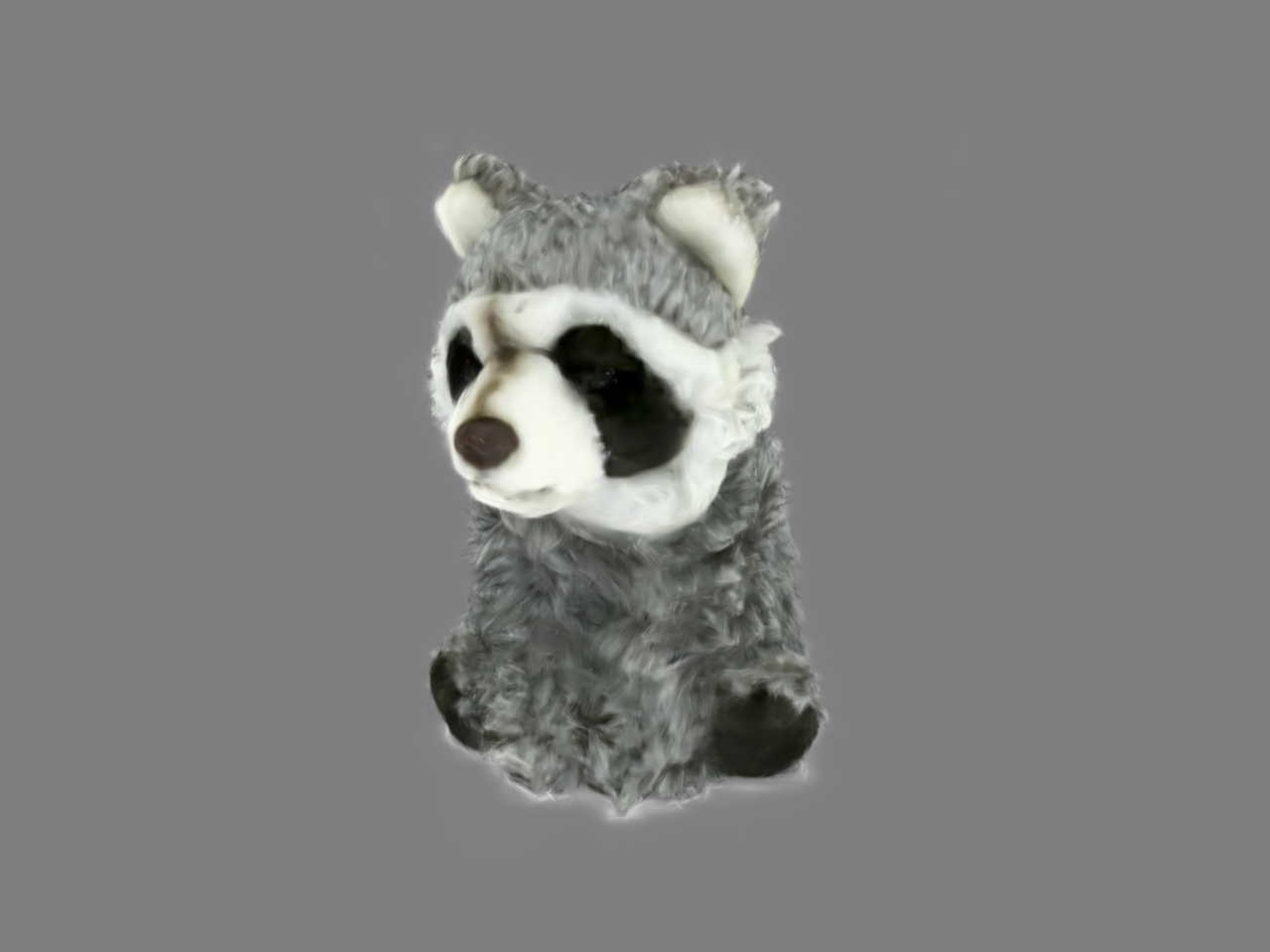}
    \\
    \vspace{-1mm}
    \makebox[0.99\columnwidth][c]{Generated assets} \\[1mm]
    \includegraphics[width=0.500\columnwidth]{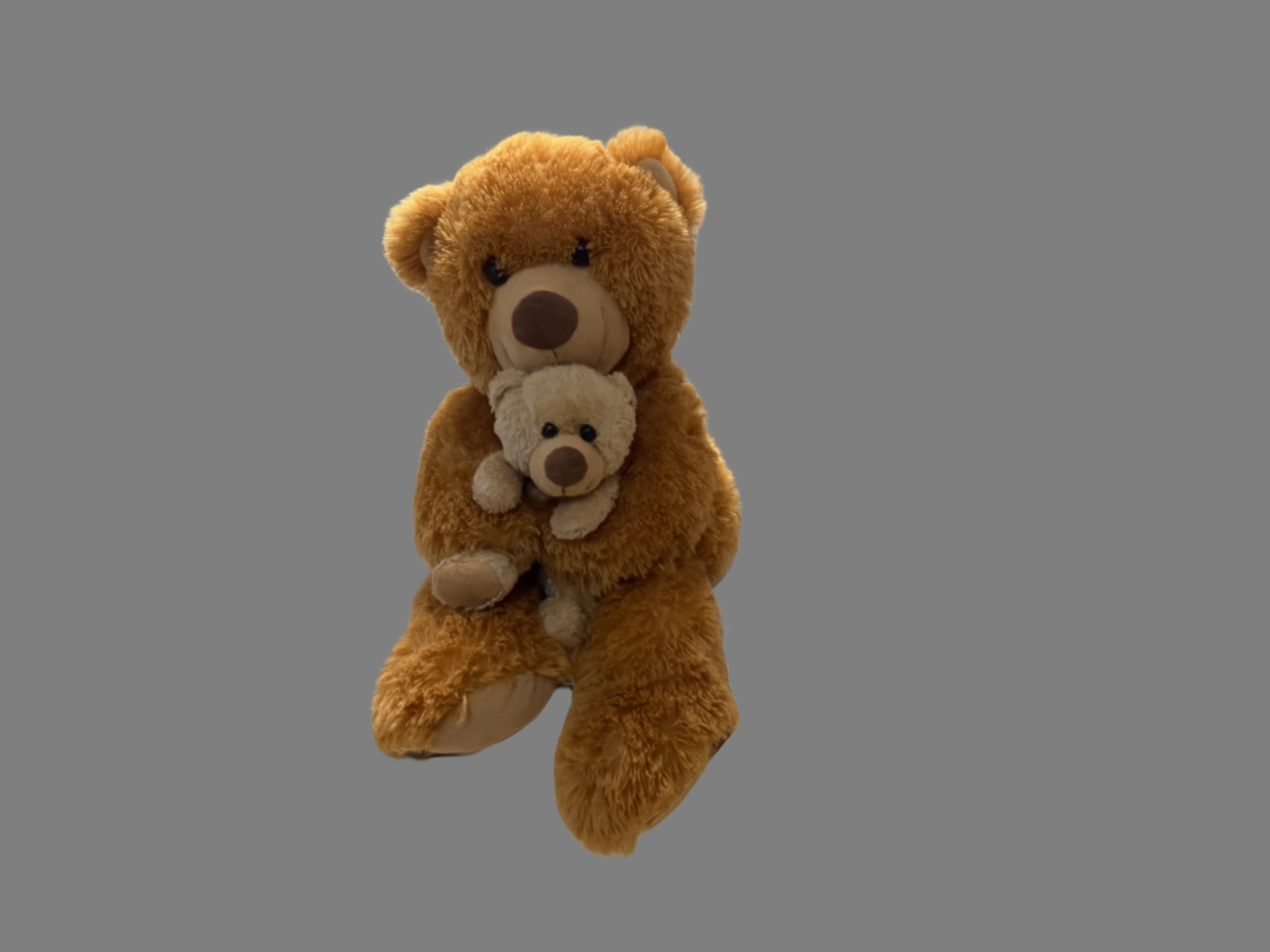}
    \includegraphics[width=0.500\columnwidth]{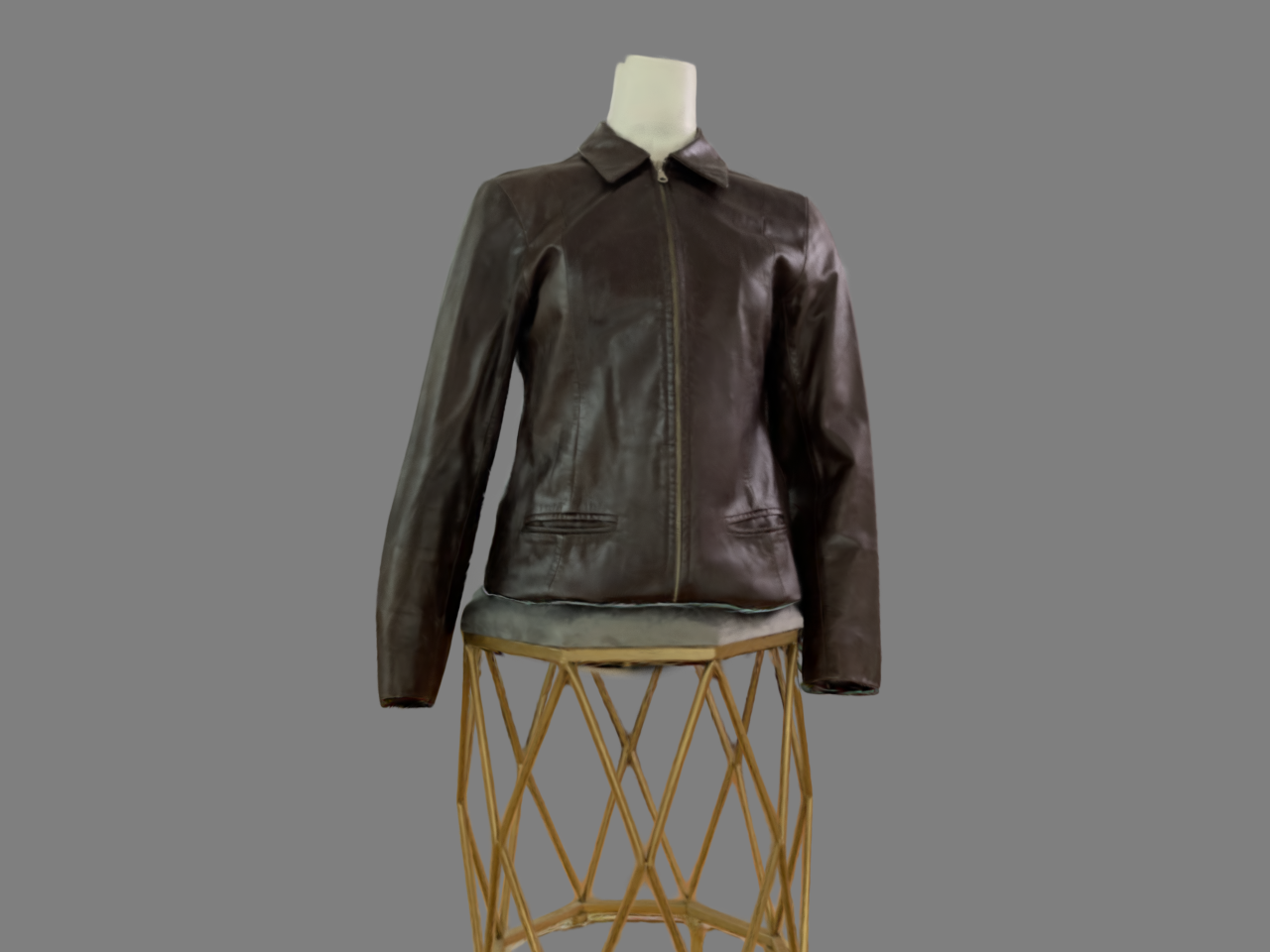}
    \includegraphics[width=0.500\columnwidth]{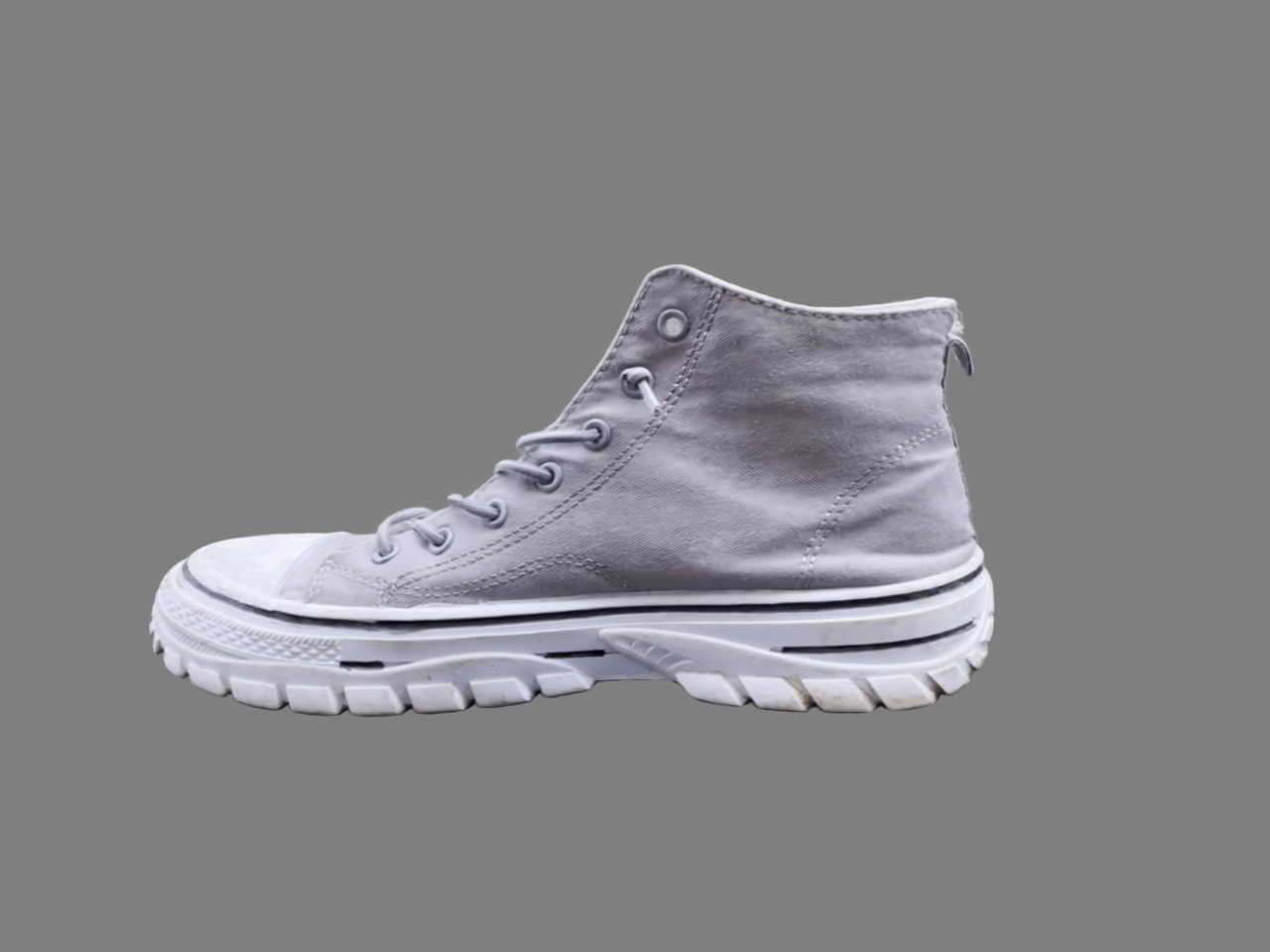}
    \includegraphics[width=0.500\columnwidth]{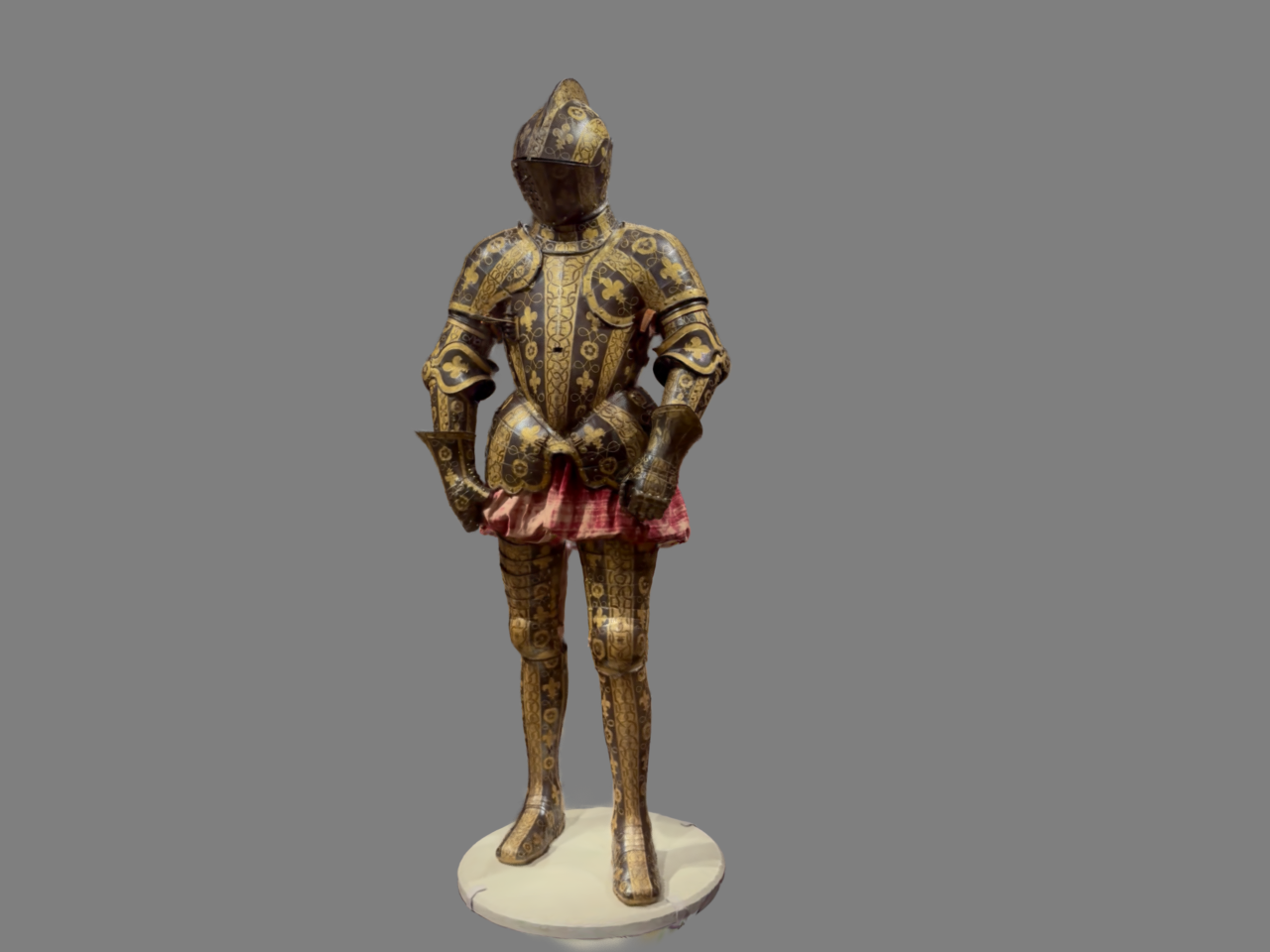}
    \\
    \vspace{-1mm}
    \makebox[0.99\columnwidth][c]{Reconstructed single-object assets} \\[1mm]
    \includegraphics[width=0.500\columnwidth]{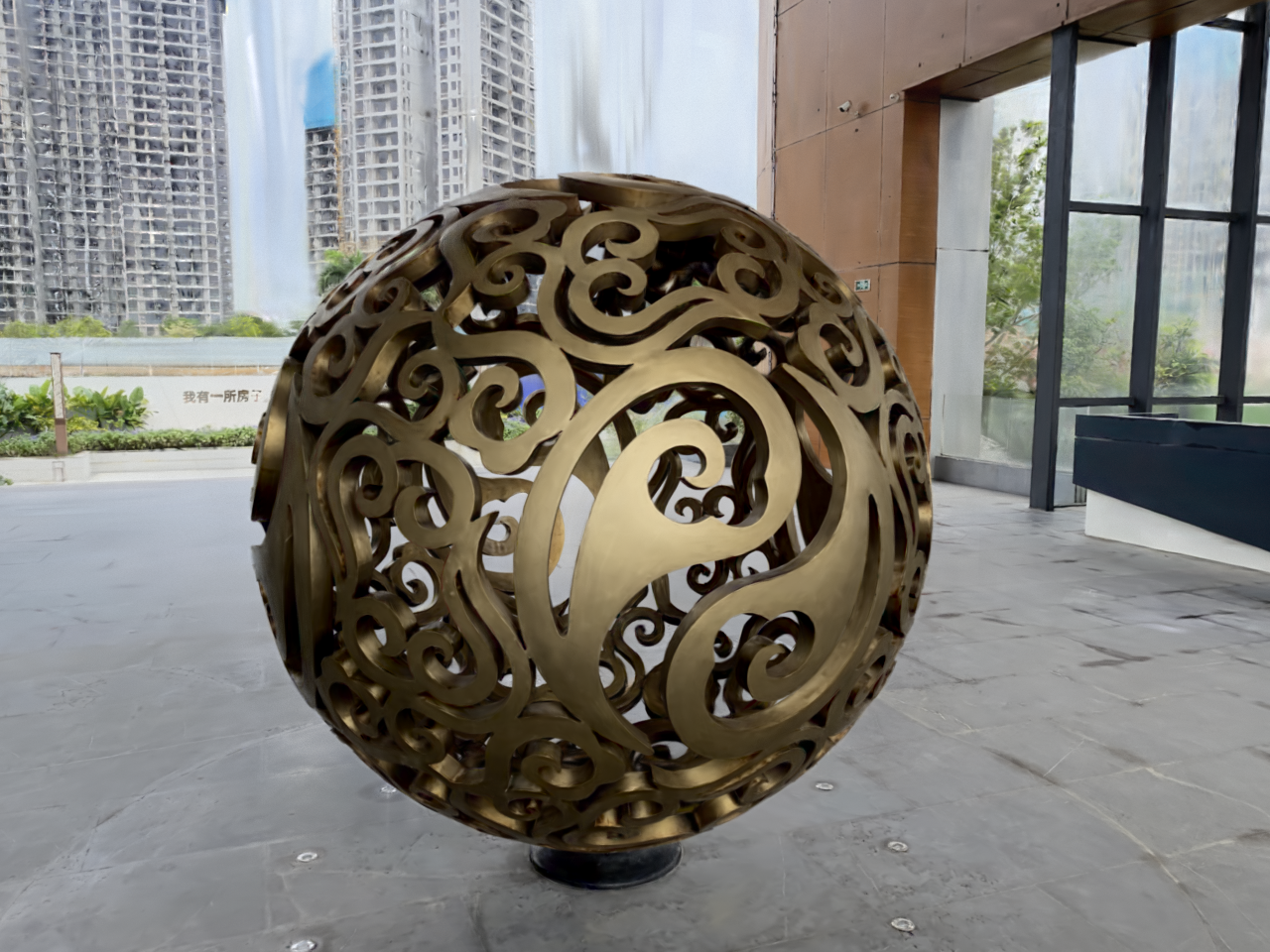}
    \includegraphics[width=0.500\columnwidth]{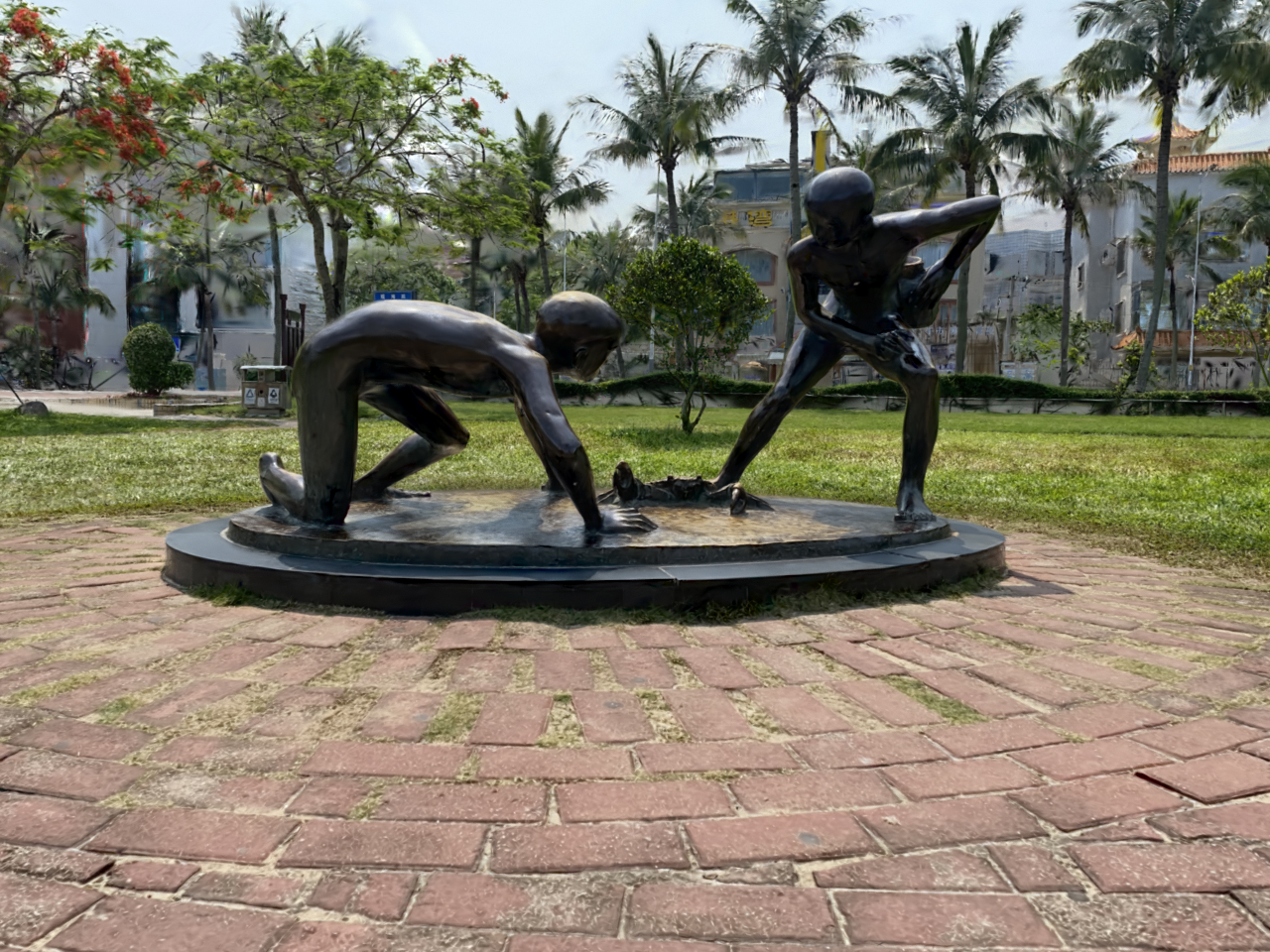}
    \includegraphics[width=0.500\columnwidth]{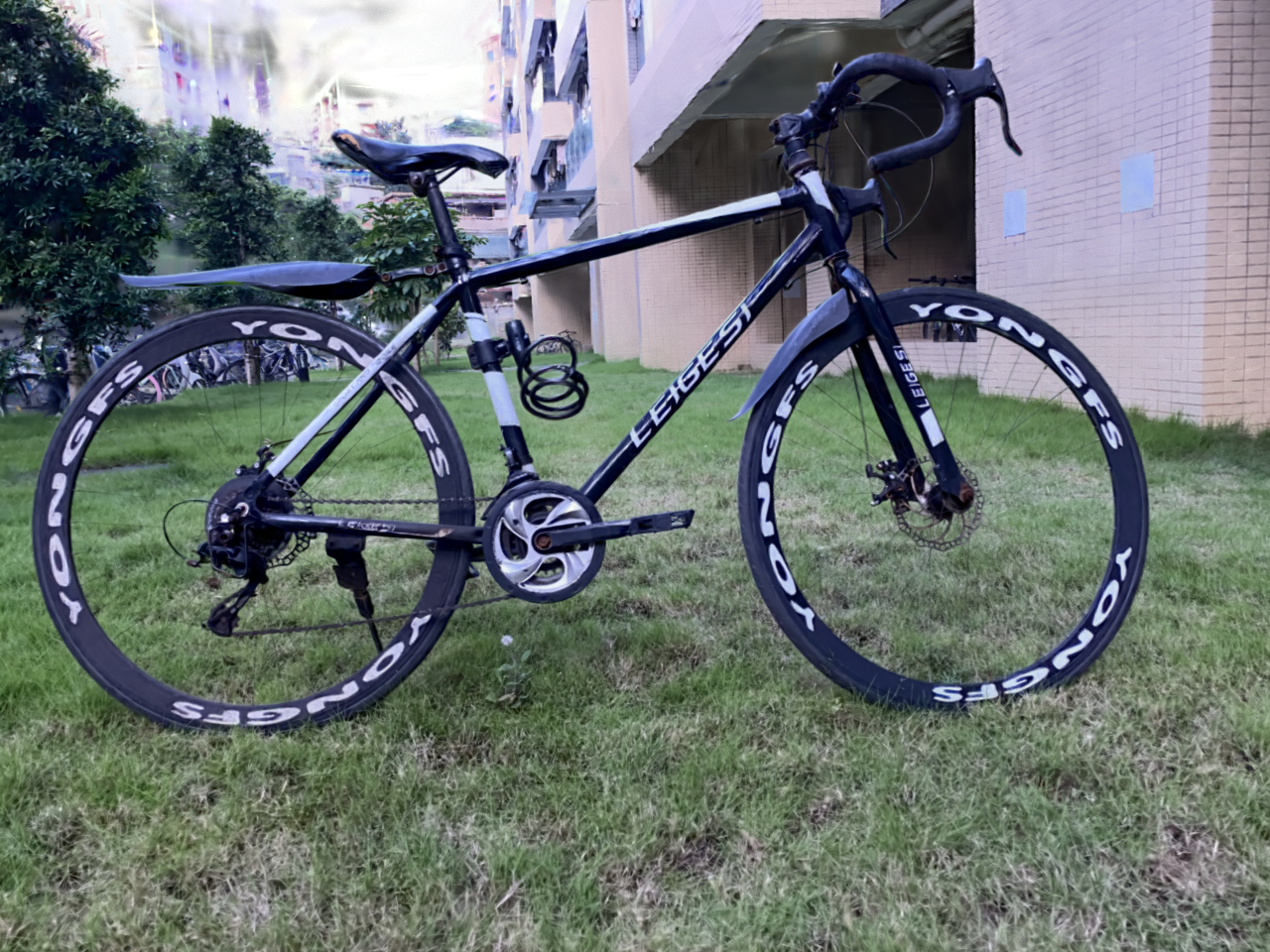}
    \includegraphics[width=0.500\columnwidth]{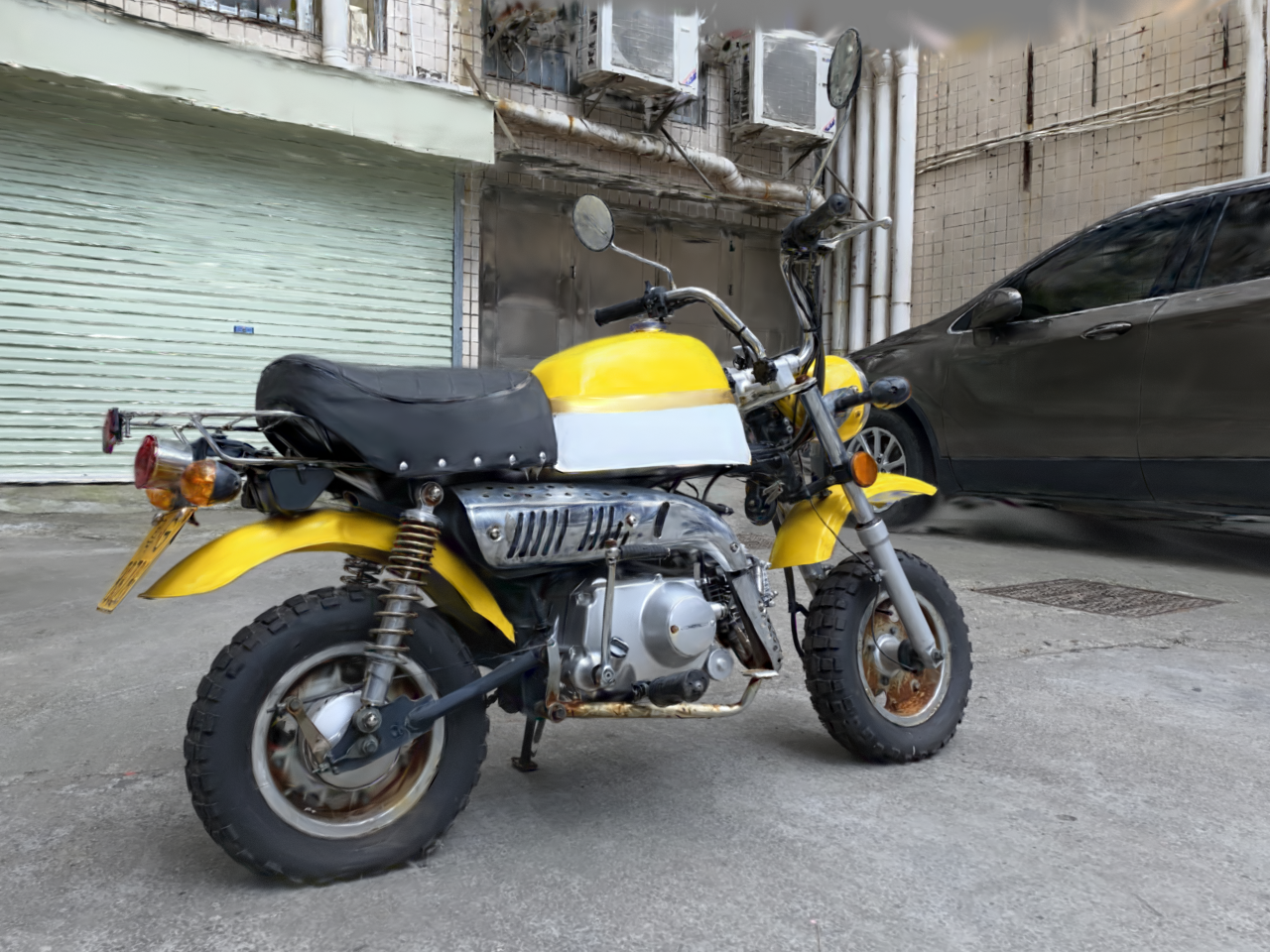}
    \\
    \vspace{-1mm}
    \makebox[0.99\columnwidth][c]{Reconstructed scenes} \\
    \caption{
        \textbf{Assets generated with different methods.}
        Our method renders well for assets from different authoring pipelines: LRM-generated, single-object reconstructed and scene-scale reconstructed assets.
    }
    \label{fig:comparison-adobe}
\end{figure*}

Our method works well with assets from different source pipelines.
In \cref{fig:comparison-adobe}, we show assets generated by large reconstruction model (LRM) \cite{gslrm} (top), single reconstructed objects (middle), and scene-scale assets (bottom).

\begin{table}
    \centering
    \small
    \caption{
        \textbf{Interactive render performance (in FPS).}
        Rasterization rendering \cite{kerbl20233dgs} is tested with their open-sourced viewer on the Windows 11 desktop.
        Rasterization outperforms our GPU ray tracer, as expected, but our method still provides a real-time experience.
        Meanwhile, our method remains interactive on low-end GPUs, even for scene assets.
    }
    \setlength\tabcolsep{1.9mm}%
    \begin{tabular}{c|ccccc}
        \toprule
        \multirow{2}{*}{\textbf{Asset}} & \multirow{2}{*}{\textbf{3DGS (Win.)}}  & \multicolumn{2}{c}{\textbf{\ouri (Win)}}      & \multicolumn{2}{c}{\textbf{\ouri (Mac)}} \\
        & & \textbf{GPU} & \textbf{CPU} & \textbf{GPU} & \textbf{CPU} \\
        \midrule
        \textit{drjohnson} & 324 & 76.4 & 1.3 & 3.6 & 0.9 \\
        \textit{playroom} & 282 & 109.9 & 2.0 & 5.5 & 1.3 \\
        \textit{room} & 314 & 108.6 & 2.0 & 5.6 & 1.3 \\
        \bottomrule
    \end{tabular}
    \label{tab:perf-interactive}
\end{table}

\paragraph*{Convergence.}

\begin{figure*}
    \centering
    \includegraphics[width=0.34\columnwidth]{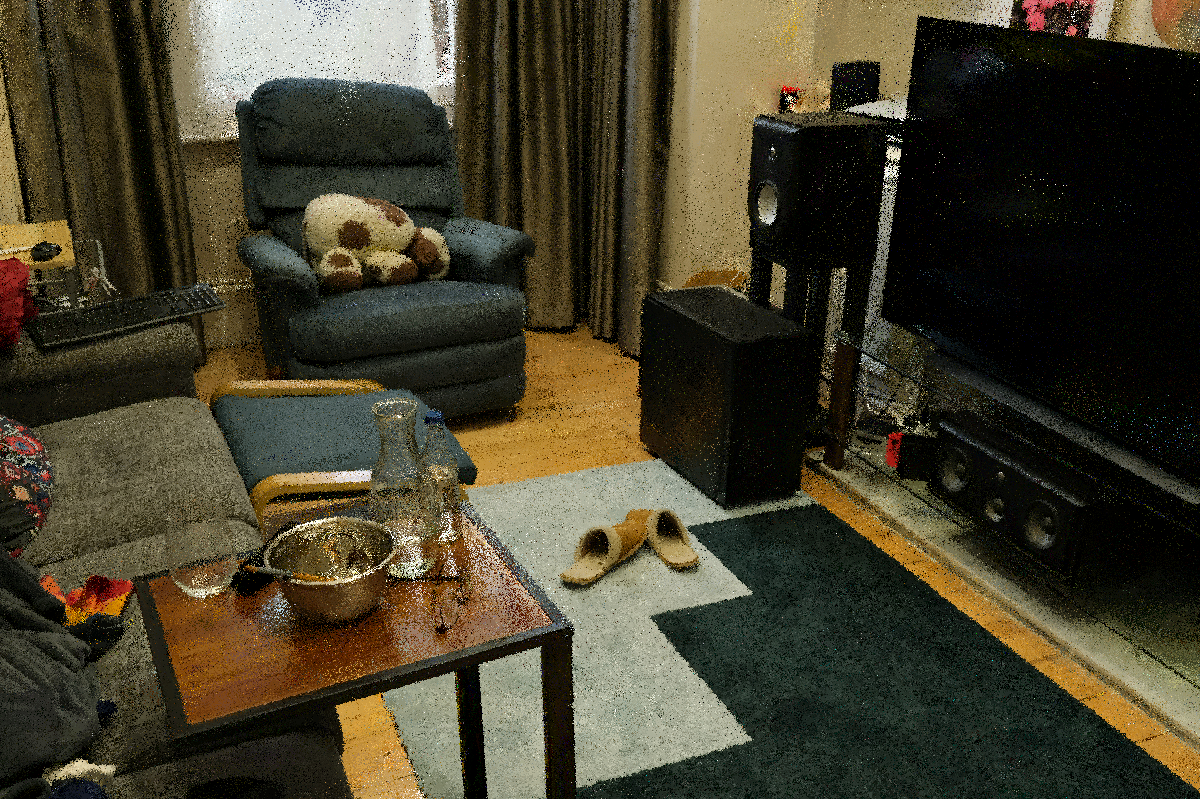}
    \includegraphics[trim={4.5in 2in 10.1667in 7.1111in},clip,width=0.17\columnwidth]{figures/comparison-inria/room-tracer-depth_intersection-spp_0001.png}
    \includegraphics[width=0.34\columnwidth]{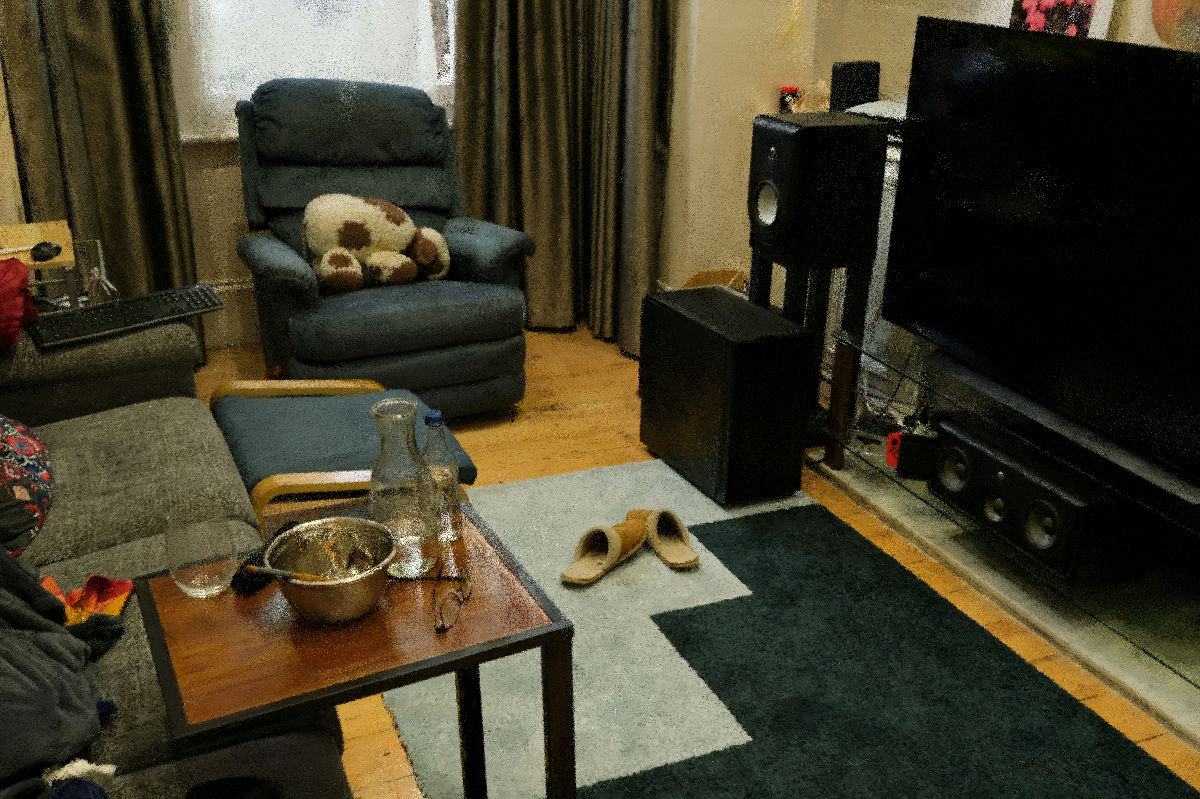}
    \includegraphics[trim={4.5in 2in 10.1667in 7.1111in},clip,width=0.17\columnwidth]{figures/comparison-inria/room-tracer-depth_intersection-spp_0004.png}
    \includegraphics[width=0.34\columnwidth]{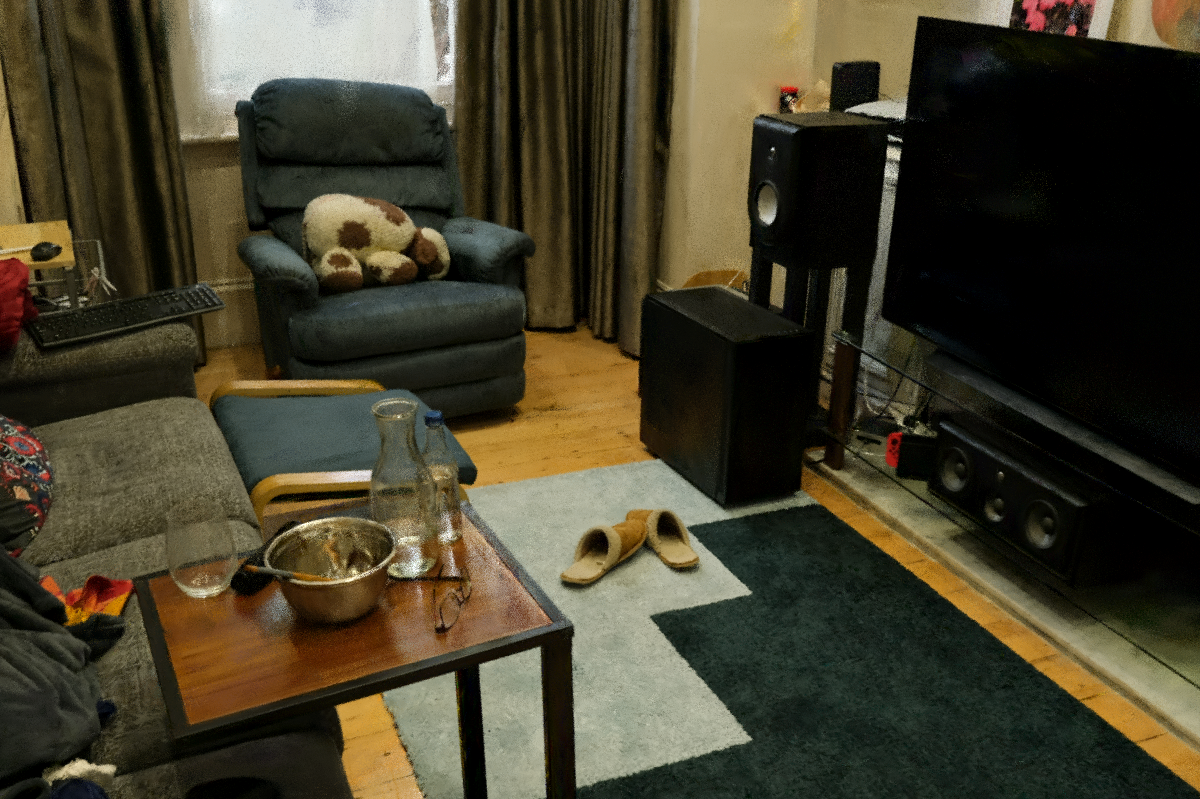}
    \includegraphics[trim={4.5in 2in 10.1667in 7.1111in},clip,width=0.17\columnwidth]{figures/comparison-inria/room-tracer-depth_intersection-spp_0016.png}
    \includegraphics[width=0.34\columnwidth]{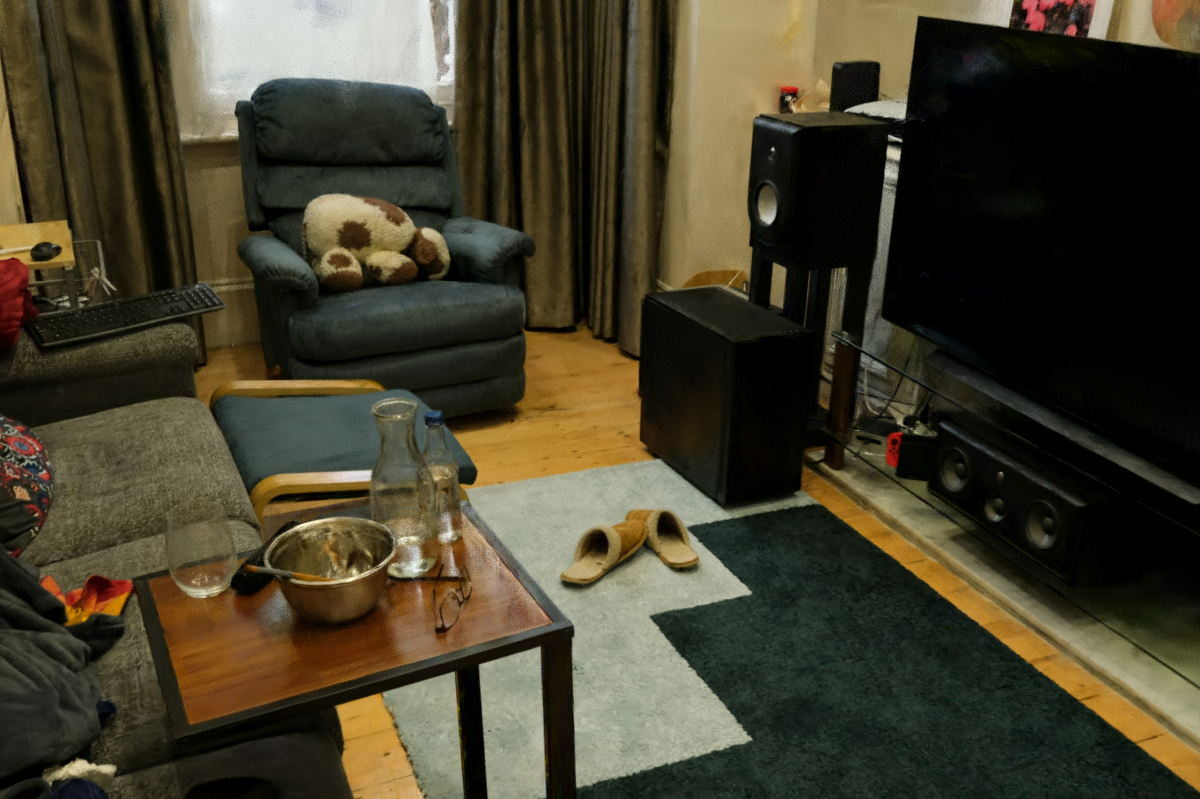}
    \includegraphics[trim={4.5in 2in 10.1667in 7.1111in},clip,width=0.17\columnwidth]{figures/comparison-inria/room-tracer-depth_intersection-spp_0064.png}
    \\
    \includegraphics[width=0.34\columnwidth]{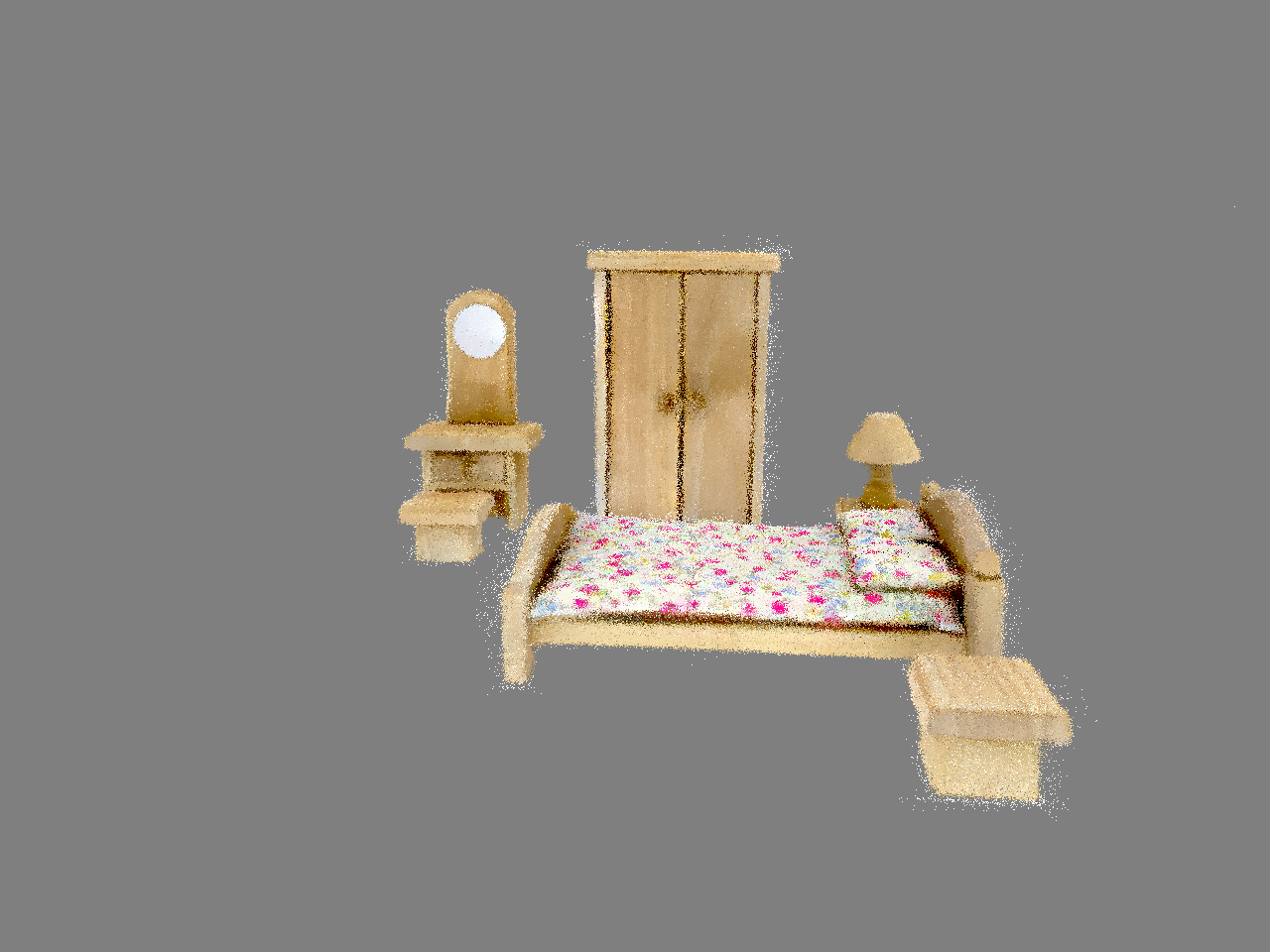}
    \includegraphics[trim={12in 4in 3.7778in 7.3333in},clip,width=0.17\columnwidth]{figures/comparison-generated-clean/generated_0063-tracer-depth_intersection-spp_0001.png}
    \includegraphics[width=0.34\columnwidth]{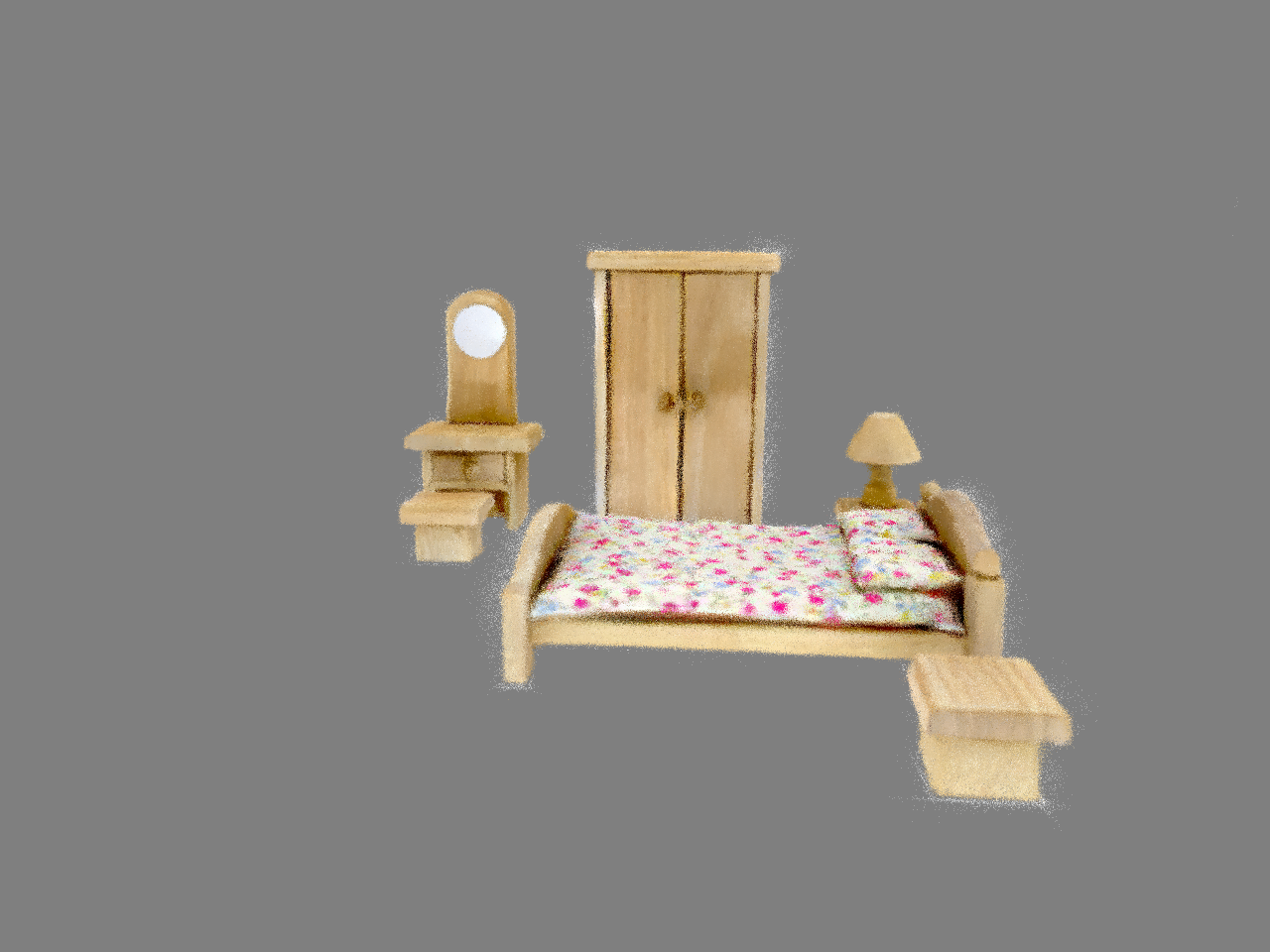}
    \includegraphics[trim={12in 4in 3.7778in 7.3333in},clip,width=0.17\columnwidth]{figures/comparison-generated-clean/generated_0063-tracer-depth_intersection-spp_0004.png}
    \includegraphics[width=0.34\columnwidth]{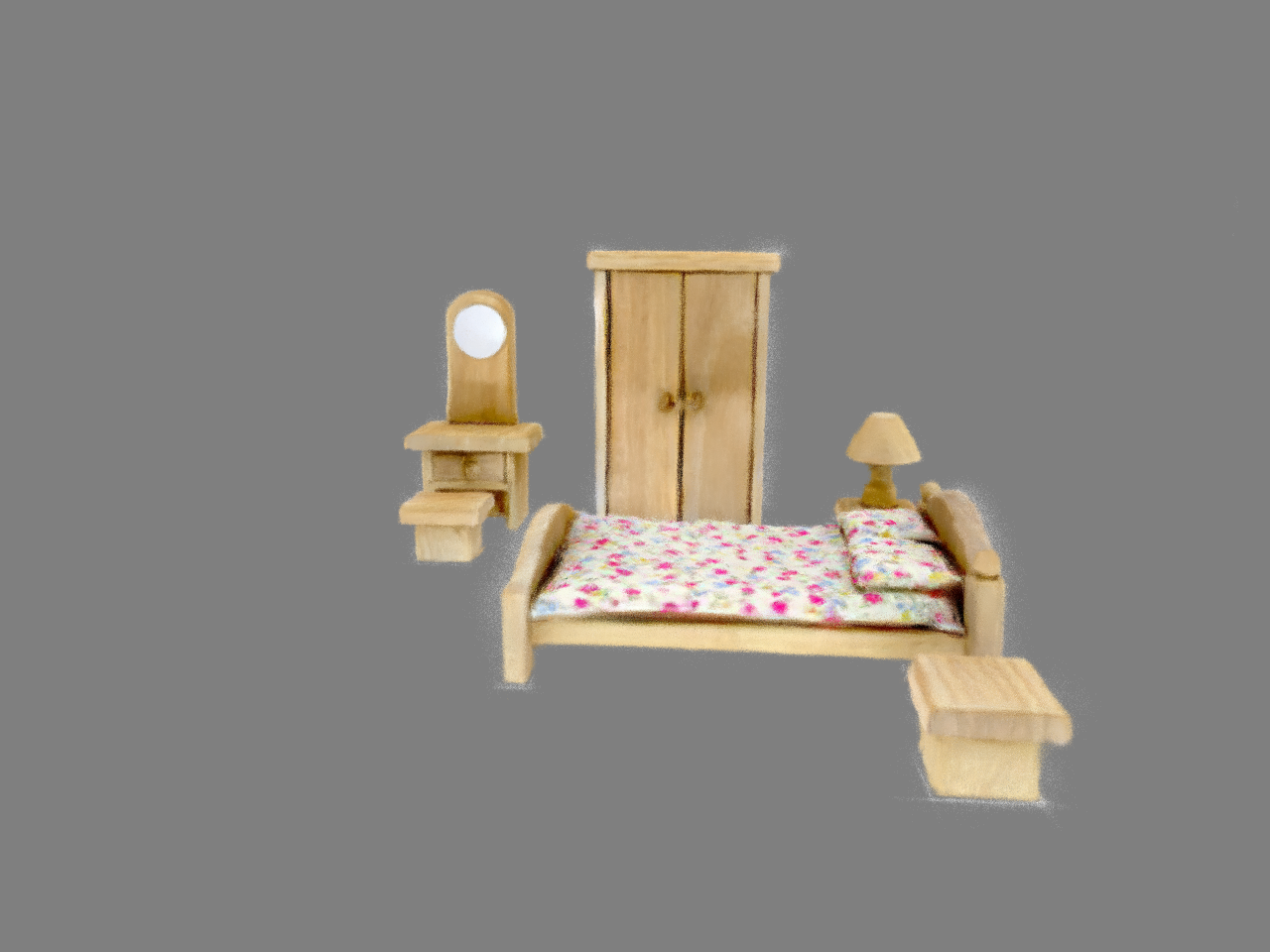}
    \includegraphics[trim={12in 4in 3.7778in 7.3333in},clip,width=0.17\columnwidth]{figures/comparison-generated-clean/generated_0063-tracer-depth_intersection-spp_0016.png}
    \includegraphics[width=0.34\columnwidth]{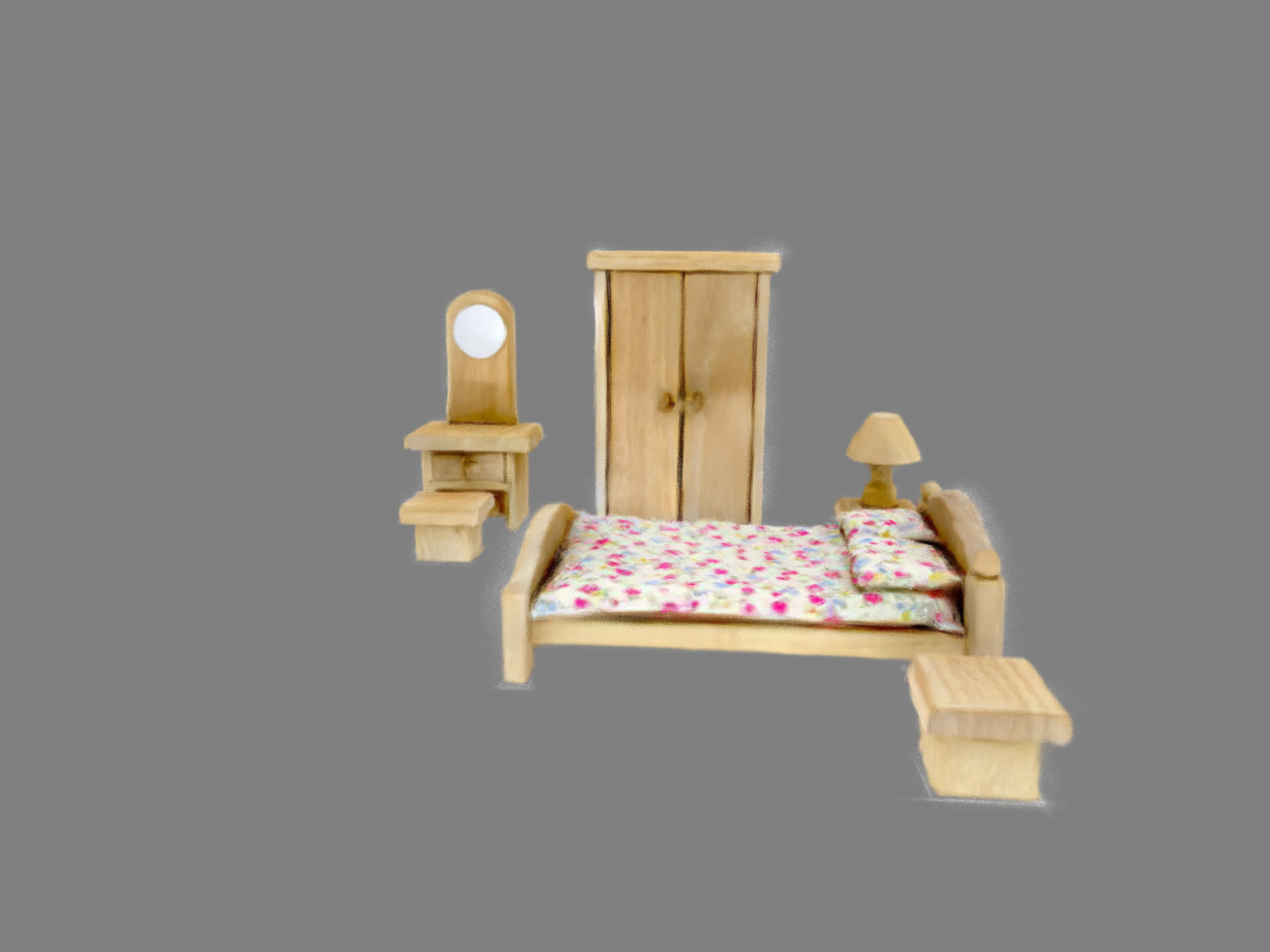}
    \includegraphics[trim={12in 4in 3.7778in 7.3333in},clip,width=0.17\columnwidth]{figures/comparison-generated-clean/generated_0063-tracer-depth_intersection-spp_0064.png}
    \\
    \includegraphics[width=0.34\columnwidth]{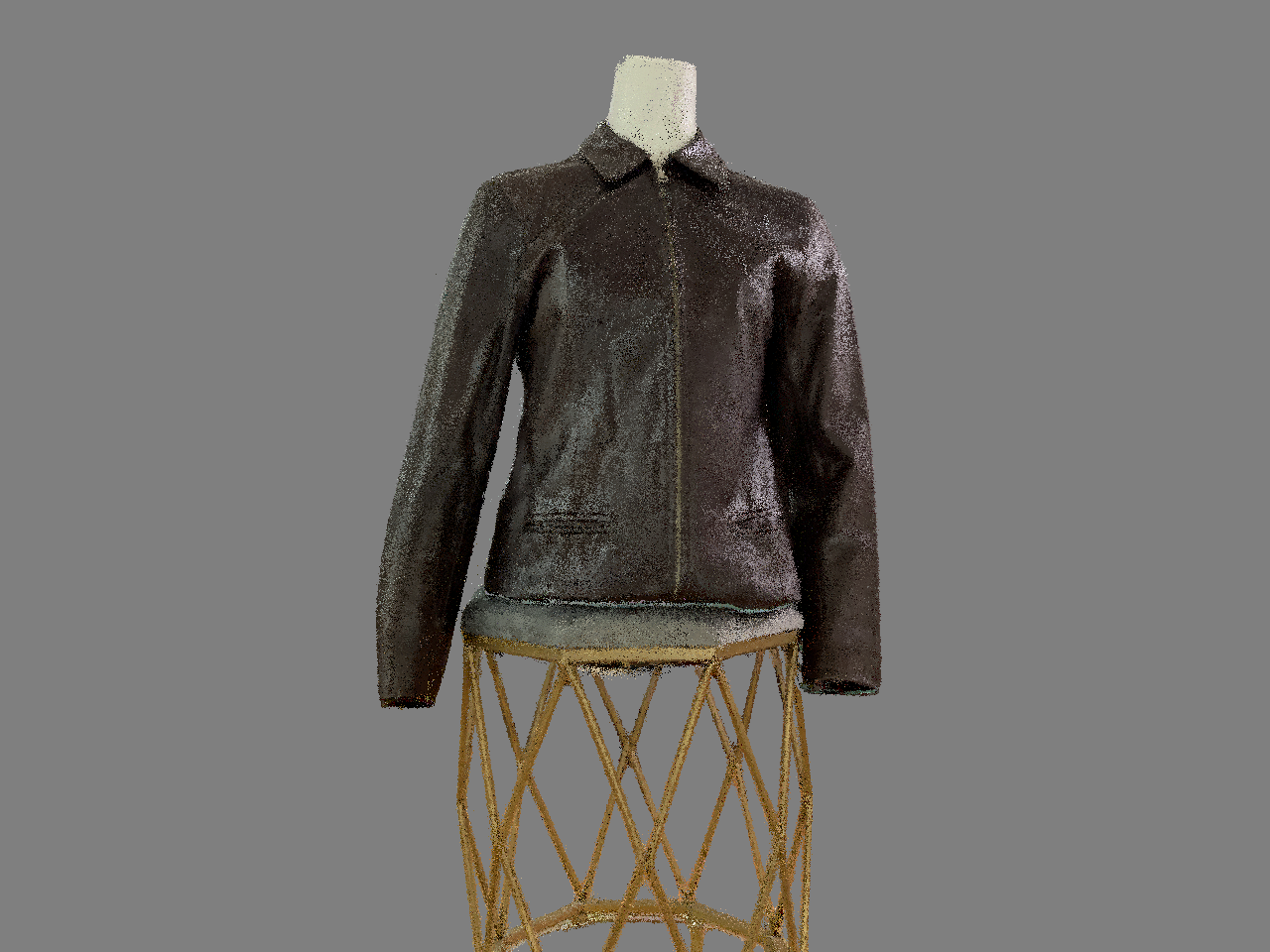}
    \includegraphics[trim={10in 8.5in 5.8778in 2.8333in},clip,width=0.17\columnwidth]{figures/comparison-reconstructed-small-clean/reconstructed_small_clean_capture_12_fashion_coat_jacket_suit_05102023-tracer-depth_intersection-spp_0001.png}
    \includegraphics[width=0.34\columnwidth]{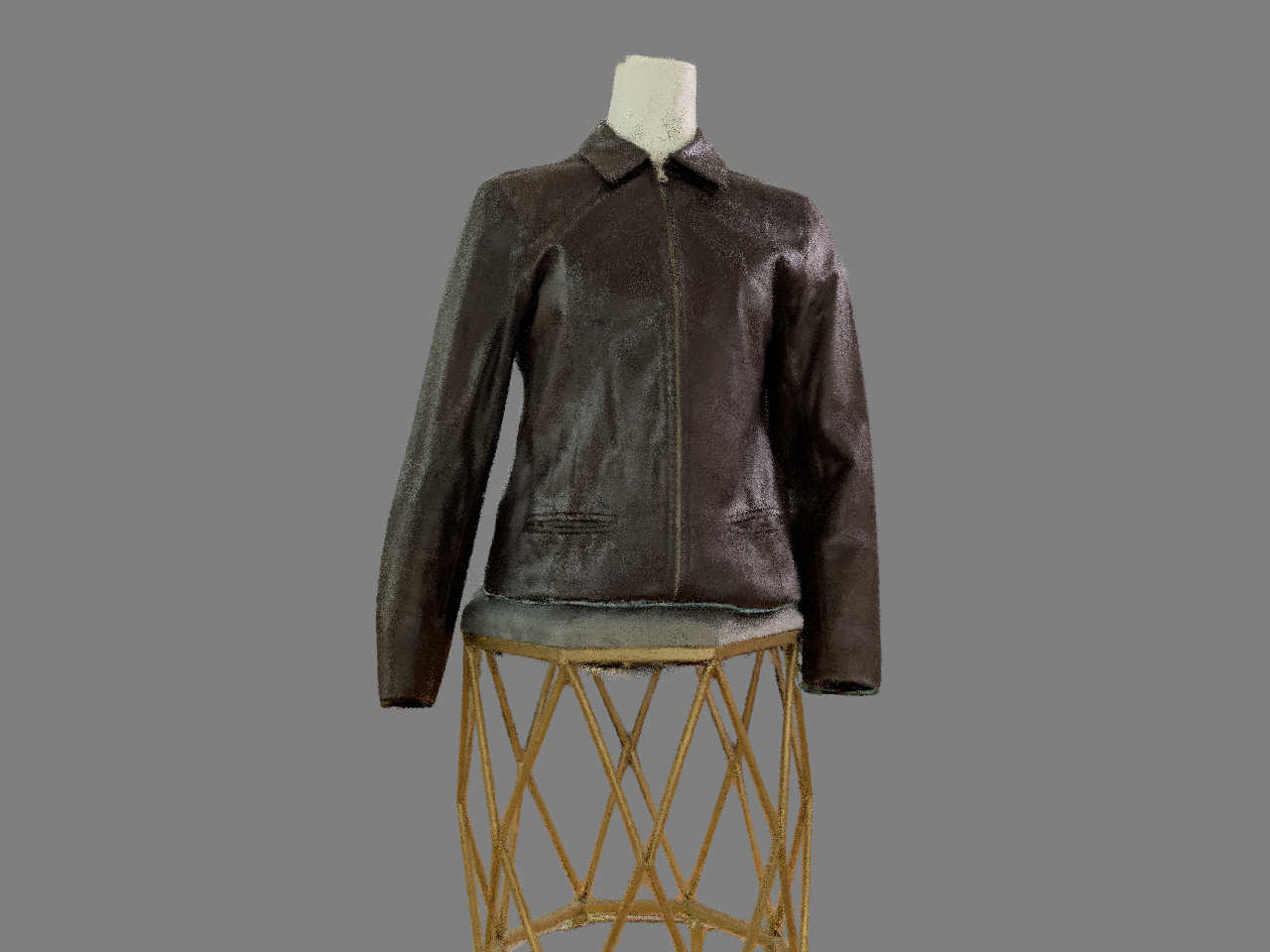}
    \includegraphics[trim={10in 8.5in 5.8778in 2.8333in},clip,width=0.17\columnwidth]{figures/comparison-reconstructed-small-clean/reconstructed_small_clean_capture_12_fashion_coat_jacket_suit_05102023-tracer-depth_intersection-spp_0004.png}
    \includegraphics[width=0.34\columnwidth]{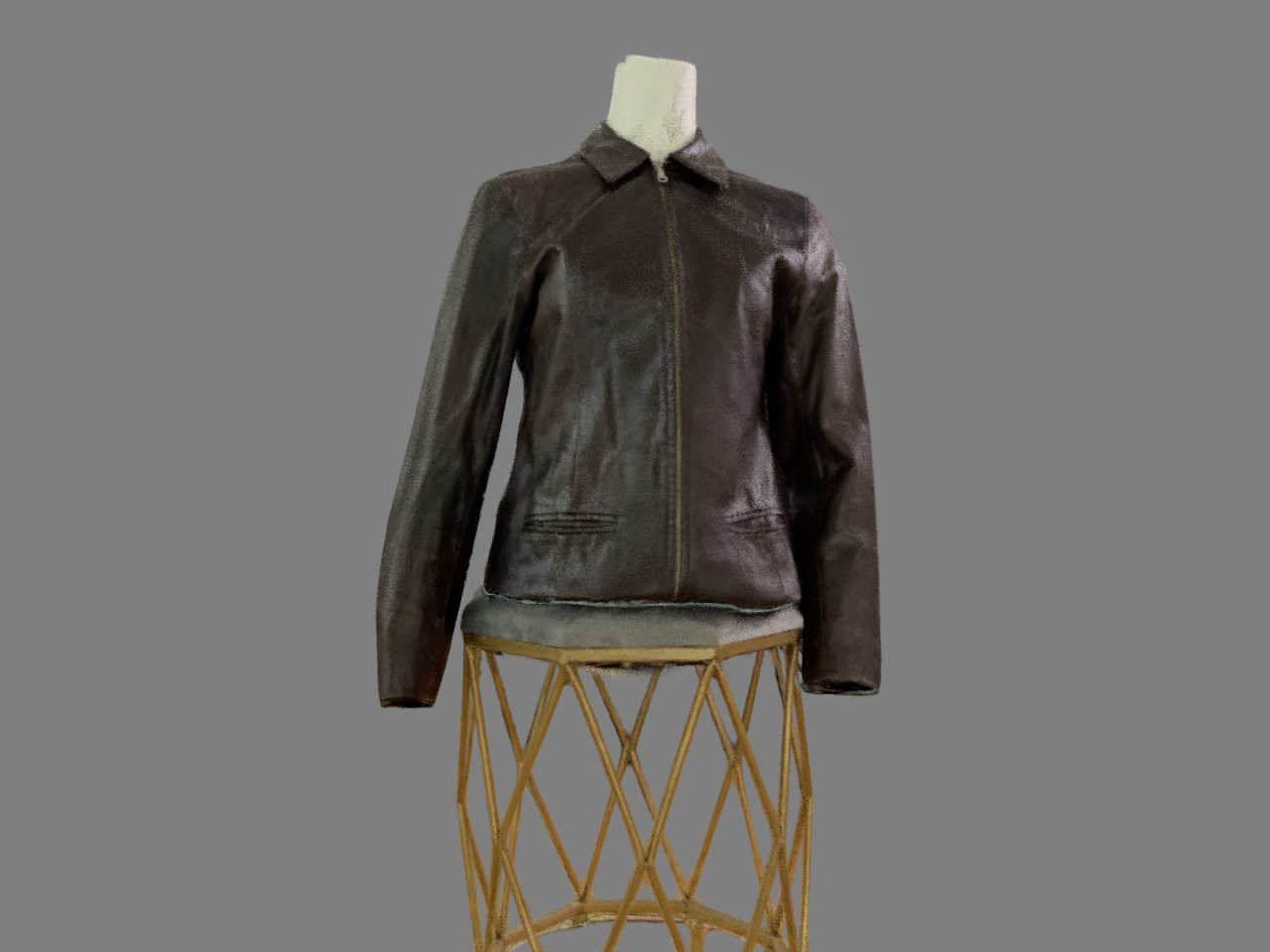}
    \includegraphics[trim={10in 8.5in 5.8778in 2.8333in},clip,width=0.17\columnwidth]{figures/comparison-reconstructed-small-clean/reconstructed_small_clean_capture_12_fashion_coat_jacket_suit_05102023-tracer-depth_intersection-spp_0016.png}
    \includegraphics[width=0.34\columnwidth]{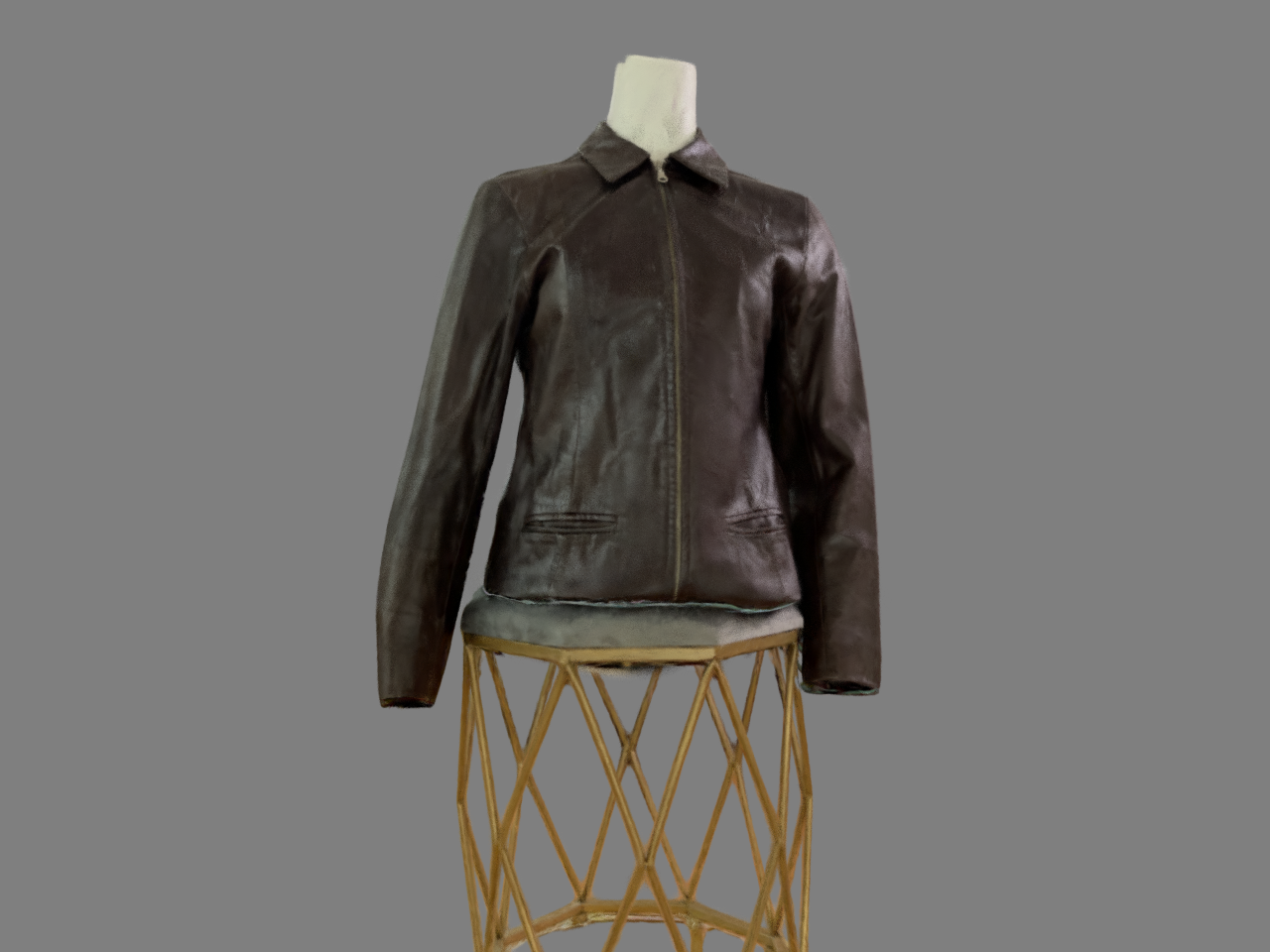}
    \includegraphics[trim={10in 8.5in 5.8778in 2.8333in},clip,width=0.17\columnwidth]{figures/comparison-reconstructed-small-clean/reconstructed_small_clean_capture_12_fashion_coat_jacket_suit_05102023-tracer-depth_intersection-spp_0064.png}
    \\
    \includegraphics[width=0.34\columnwidth]{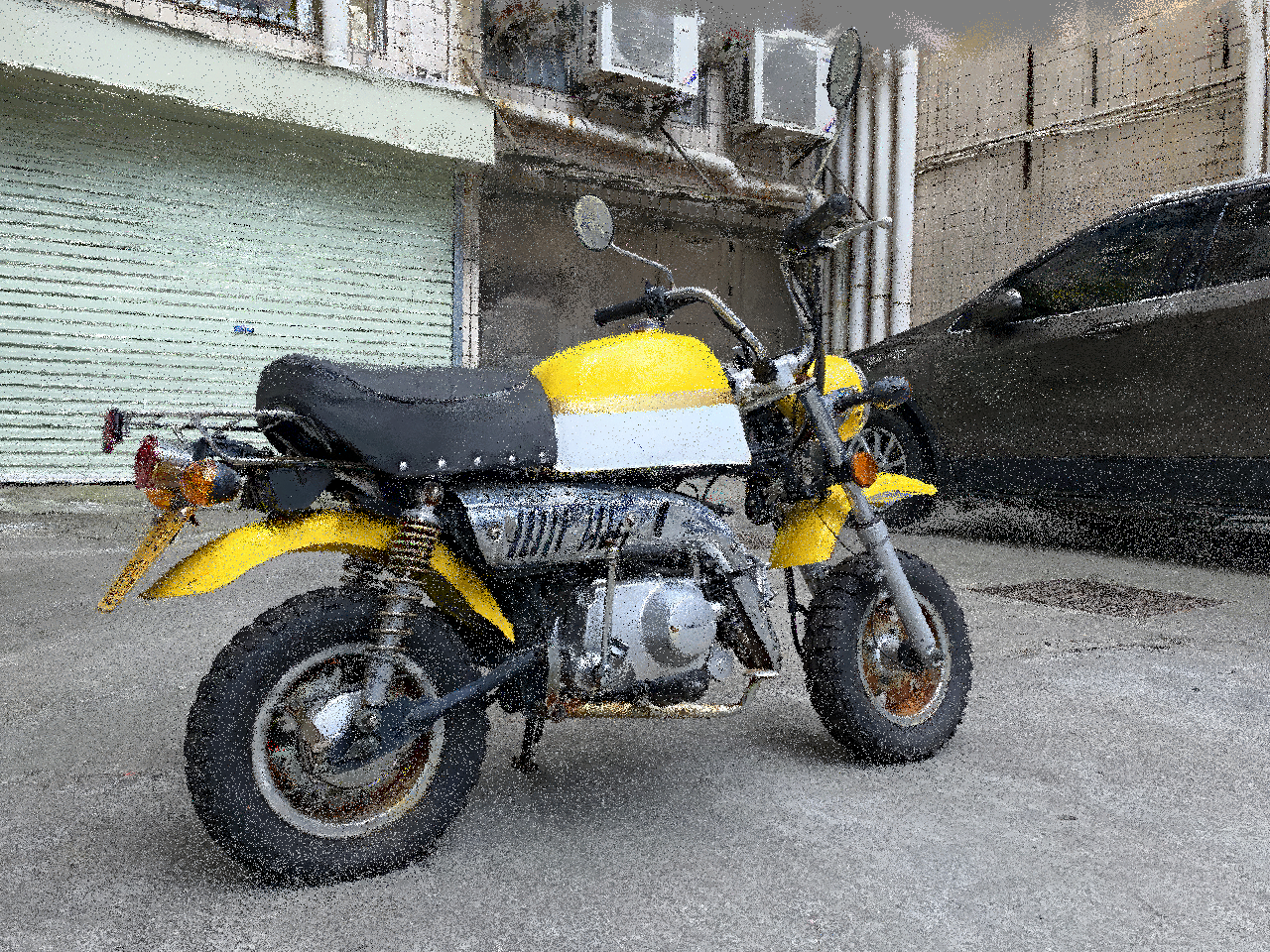}
    \includegraphics[trim={5.5in 4.5in 10.2778in 6.8333in},clip,width=0.17\columnwidth]{figures/comparison-reconstructed-large/reconstructed_large_0190-tracer-depth_intersection-spp_0001.png}
    \includegraphics[width=0.34\columnwidth]{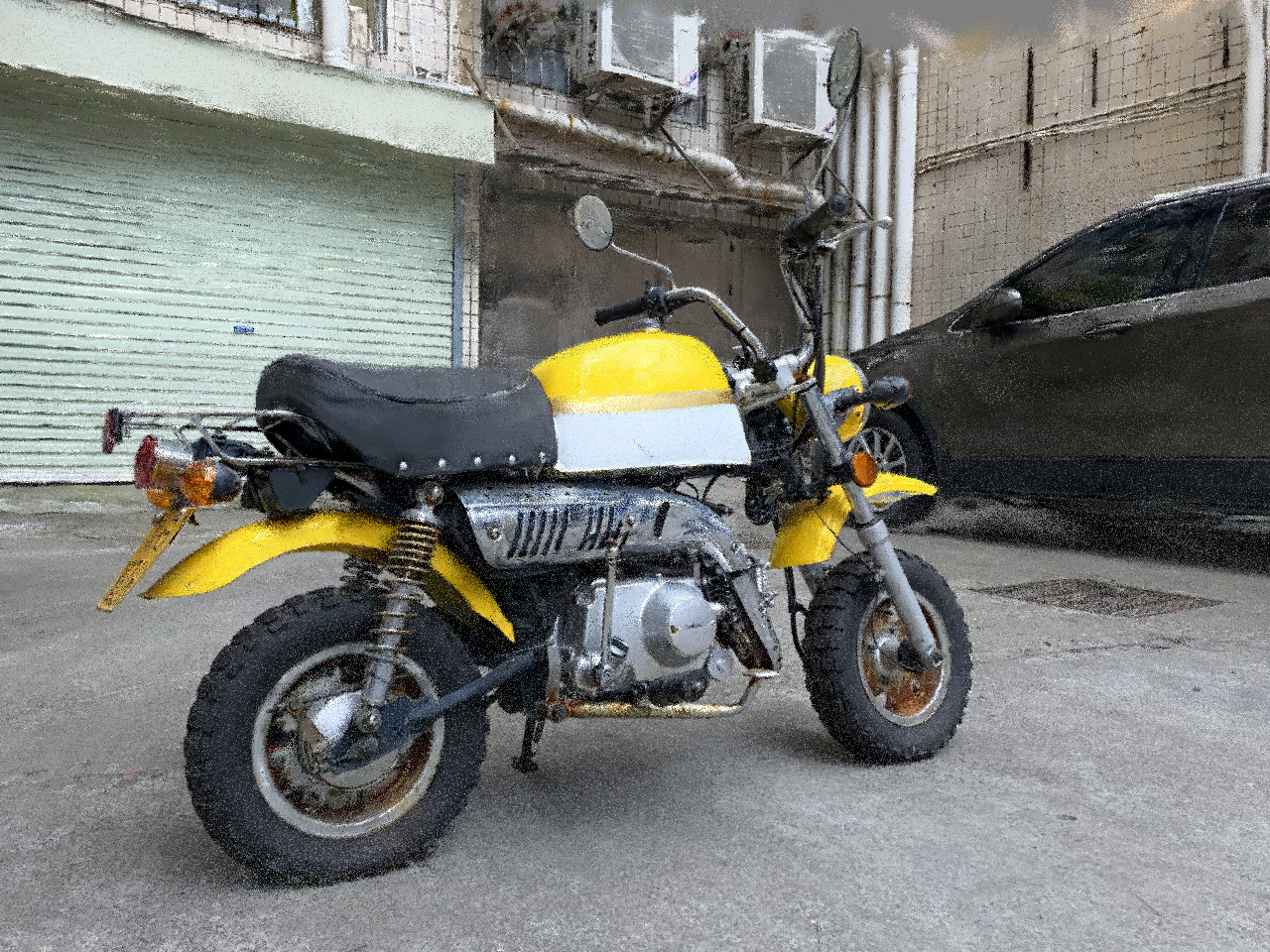}
    \includegraphics[trim={5.5in 4.5in 10.2778in 6.8333in},clip,width=0.17\columnwidth]{figures/comparison-reconstructed-large/reconstructed_large_0190-tracer-depth_intersection-spp_0004.png}
    \includegraphics[width=0.34\columnwidth]{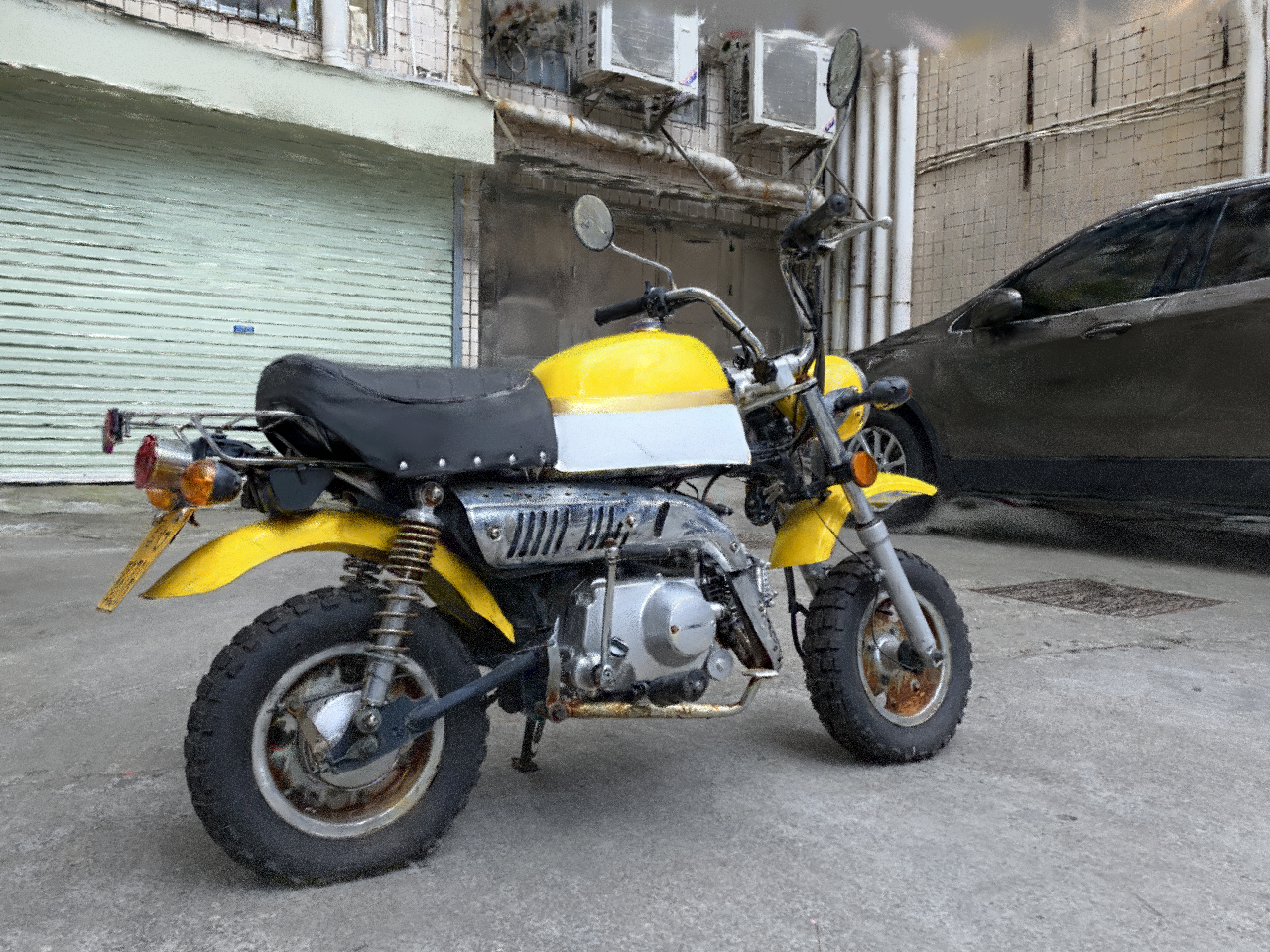}
    \includegraphics[trim={5.5in 4.5in 10.2778in 6.8333in},clip,width=0.17\columnwidth]{figures/comparison-reconstructed-large/reconstructed_large_0190-tracer-depth_intersection-spp_0016.png}
    \includegraphics[width=0.34\columnwidth]{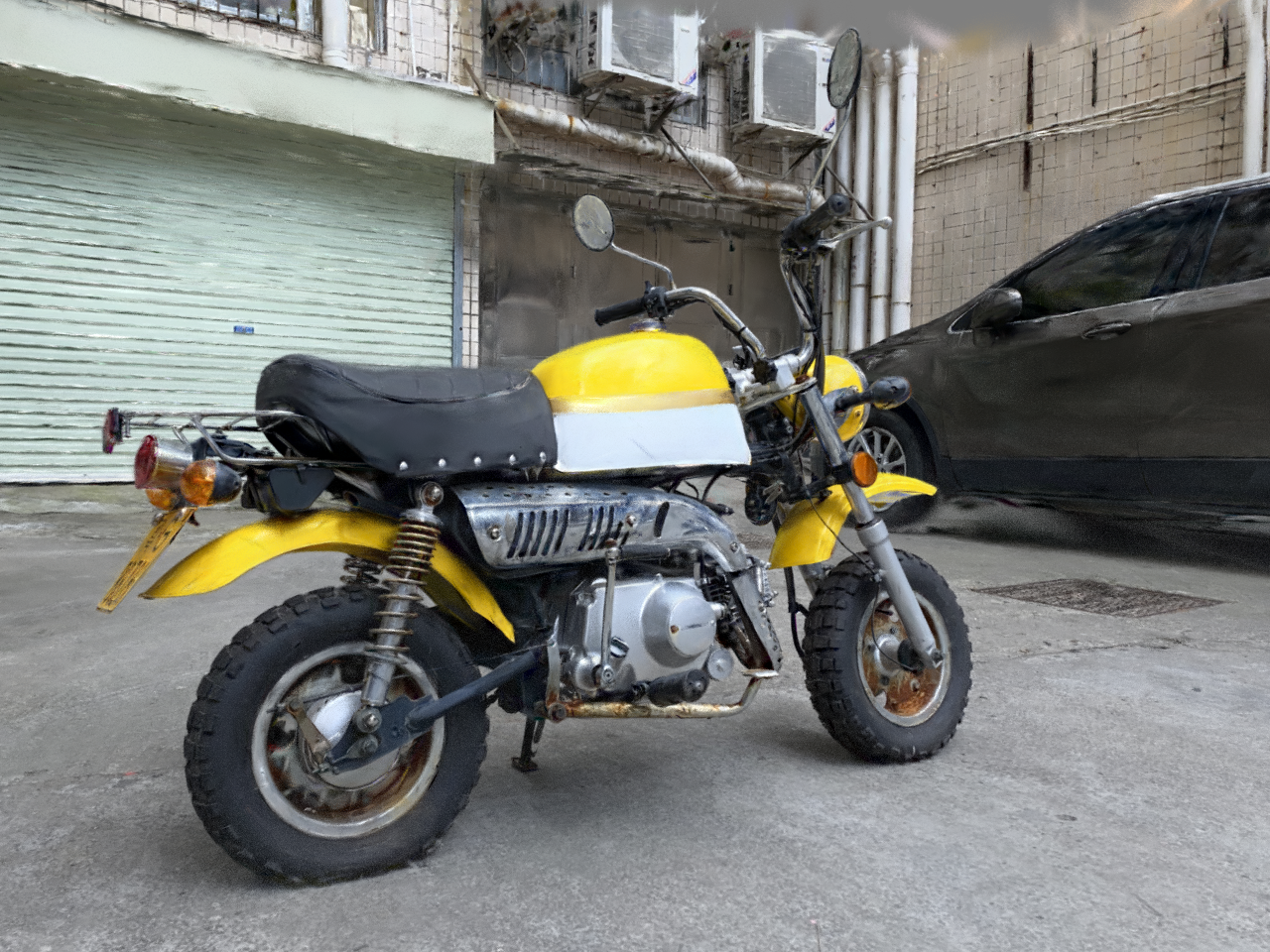}
    \includegraphics[trim={5.5in 4.5in 10.2778in 6.8333in},clip,width=0.17\columnwidth]{figures/comparison-reconstructed-large/reconstructed_large_0190-tracer-depth_intersection-spp_0064.png}
    \\
    \makebox[0.518\columnwidth][c]{1\,spp}
    \makebox[0.518\columnwidth][c]{4\,spp}
    \makebox[0.518\columnwidth][c]{16\,spp}
    \makebox[0.518\columnwidth][c]{64\,spp}
    \vspace{-2mm}
    \caption{
        \textbf{Convergence of our method.}
        Although our stochastic binary opacity introduces Monte Carlo noise, it converges very fast.
        For all assets we tested, 1 spp already produces plausible rendering, and most noise is eliminated with 64 spp or less.
    }
    \label{fig:comparison-convergence}
\end{figure*}

\begin{figure*}
    \centering
    \includegraphics[width=\textwidth]{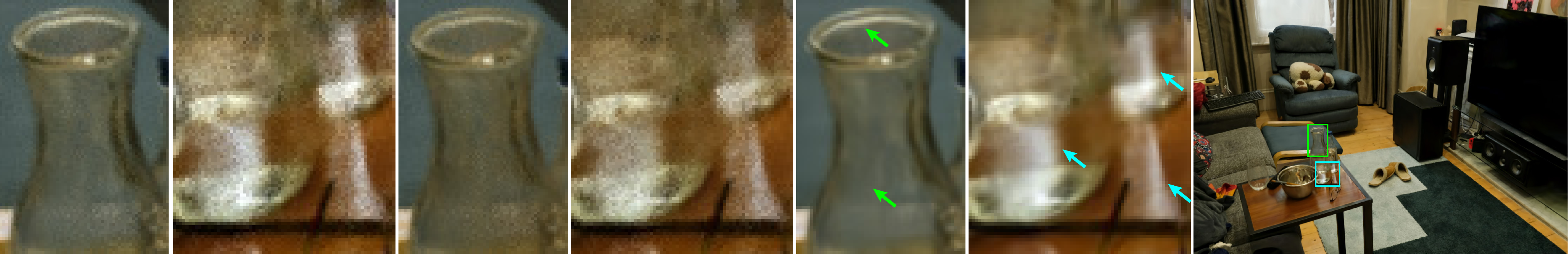}\\
    \makebox[0.52\columnwidth][c]{\textbf{Ours}}
    \makebox[0.52\columnwidth][c]{Stochastic depth sampling}
    \makebox[0.52\columnwidth][c]{Reference (zoom-in)}
    \makebox[0.47\columnwidth][c]{Reference (full)}
    \caption{
        \textbf{Bias from stochastic depth sampling in 3DGRT.}
        The stochastic depth sampling in 3DGRT is a biased approximation.
        We implemented it in our renderer for comparison since the official 3DGRT implementation does not support it.
        The results are rendered with 64\,spp. The bias of stochastic depth sampling becomes obvious when setting $k=1$.
        In the left zoom-in, the cloth color through the glass becomes more yellow.
        In the right zoom-in, the caustics from the refractions and reflections are dimmer.
    }
    \label{fig:comparison-bias}
\end{figure*}

While the stochastic binary opacity \cref{sec:stochastic:opacity} introduces noise, 1 spp already produces reasonable renderings, and most noise is eliminated with 64 spp or less (\cref{fig:comparison-convergence}).
While our method renders an unbiased estimate of the radiance, the stochastic depth sampling in 3DGRT is a biased approximation. If an intersected Gaussian is sampled with probability matching the opacity of the intersection, it must be treated as fully opaque for unbiased shading (technically, division by the probability cancels out the opacity). In 3DGRT, stochastic depth sampling collects the first $k$ accepted transparent intersections, and their order is preserved, resulting in a biased approximation. The bias becomes more pronounced when $k$ is small and intersections have low opacity. Thus, $k$ cannot be practically reduced to $1$. A comparison showing the bias is shown in \cref{fig:comparison-bias}, based on a close-up view of the room scene from \cref{fig:comparison-inria}.

\paragraph*{Multi-sample rendering.}

In \cref{tab:mstiming,tab:mstiming-windows}, we compare performance under different multi-sampling settings (i.e., the number $N$ of samples taken in a single BVH traversal) while maintaining equivalent output quality. Performance improves as more samples are traced per pass, provided that parallelism is not significantly compromised.
This trend holds for the Apple Silicon backends, which benefit from a sufficiently large L1 cache, and CPU backends. Both also inherently offer limited parallelism due to the lower number of cores. Instead, on the NVIDIA GPUs with many more cores, the performance increases up to 8$\times$ multi-sampling on NVIDIA GPUs, but significantly degrades beyond that point. The payloads for all the rays share the L1 cache, and a larger payload means less parallelism and a lower computational occupancy. In addition, it can also be partly attributed to our integration within a feature-rich production renderer---even without Gaussians the ray payload is already 244 bytes.
Our method adds no extra payload unless multi-sampling is enabled, highlighting its practicality in real-world contexts where algorithms must compete for limited computing and memory resources alongside other system components.
If we customize the renderer by disabling advanced effects, reducing the ray payload to just $56$ bytes, it will bring a further $10\%$ speedup on Windows with RTX3090 and $25\%$ speedup on Apple M1 Max.

Implementing multi-sampling is further constrained by some graphics APIs (e.g., Vulkan and DirectX Raytracing). These APIs restrict payload access to any-hit shaders, while the ray’s maximum distance ($t_\mathrm{max}$) can only be modified from intersection shaders. Nevertheless, the maximum distance can only be computed from multiple samples stored in the ray payload. Due to such a dilemma, we have to report the maximal possible hit distance in the intersection shader, and rays cannot be shortened once intersections are accepted, leading to more BVH traversal and associated performance loss. This effect is evident when comparing \cref{tab:perf-offline-windows} with the $1024\times1$ case in \cref{tab:mstiming-windows}.
While multi-sampling can effectively boost performance, our method with a single sample is a more general and robust solution for feature-rich renderers.

\begin{table}[t]
    \centering
    \small
    \caption{
        \textbf{Timing of different multi-sampling settings on MacOS, measured in seconds.} Here $m\times n$ means $m$ passes are used and each pass performs $n$-multi-sampling in the ray payload. Acceleration over using the single sampling (i.e., $1024\times 1$) ranges from $2.9\times$ to $9.5\times$ on with a Metal GPU pathtracer, and ranges from $2.5\times$ to $4.3\times$ on the CPU. The minimal timings for each scene or device are marked in bold.
    }
    \setlength\tabcolsep{2.5mm}%
    \begin{tabular}{c|cccccc}
        \toprule
                      & \multicolumn{2}{c}{\textit{drjohnson}} & \multicolumn{2}{c}{\textit{playroom}} & \multicolumn{2}{c}{\textit{room}} \\
                      & \textbf{GPU}       & \textbf{CPU}        & \textbf{GPU}         & \textbf{CPU}          & \textbf{GPU}         & \textbf{CPU}           \\
        \midrule
        1024$\times$1 & 276.7         & 1101.9          & 179.8         & 747.3             & 175.4             & 771.7             \\
        256$\times$4  & 97.8          & 467.8           & 63.4          & 304.8             & 60.3              & 281.5             \\
        64$\times$16  & 44.1          & 330.2           & 29.8          & 212.7             & 28.0              & 177.1             \\
        16$\times$64  & 30.6          & \textbf{302.1}  & 22.6          & \textbf{185.0}    & 21.6              & \textbf{149.1}    \\
        4$\times$256  & \textbf{26.9} & 304.9           & \textbf{19.7} & 188.0             & \textbf{19.4}     & 152.0             \\
        \bottomrule
    \end{tabular}
    \label{tab:mstiming}
\end{table}

\begin{figure}[t]
    \includegraphics[width=1.0\linewidth]{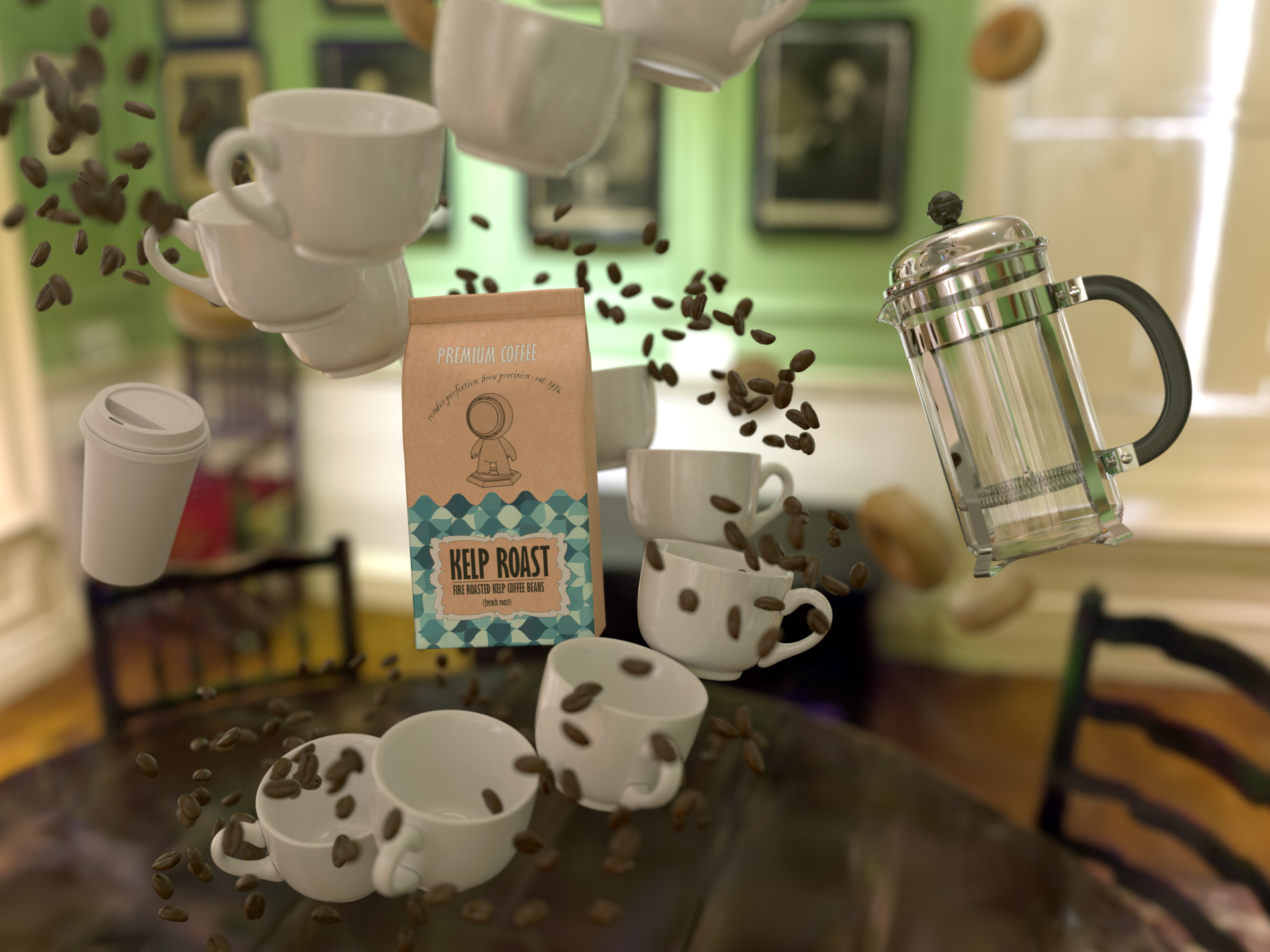}
    \caption{
        \textbf{Dr. Johnson's Coffee Whirl.}
        We mix assets made of meshes and complex materials within the \emph{drjohnson} Gaussian splatting scene asset. See also videos in supplementary materials.
    }
    \label{fig:drjohnsoncafe}
\end{figure}

\begin{table}[t]
    \centering
    \small
    \caption{
        \textbf{Timing of different multi-sampling settings on Windows, measured in seconds.} Here $m\times n$ again means $m$ passes are used and each pass performs $n$-multi-sampling in the ray payload.
        For the NVIDIA GPU, multi-sampling of 8 is the most optimal, about $4.5\times$ faster compared with $1024 \times 1$.
        Higher multi-sampling gets inferior performance due to low parallelism caused by the increasing sizes of payloads.
        The CPU implementation persistently shows better performance with higher multi-sampling.
    }
    \setlength\tabcolsep{2mm}%
    \begin{tabular}{c|cccccc}
        \toprule
                      & \multicolumn{2}{c}{\textit{drjohnson}} & \multicolumn{2}{c}{\textit{playroom}} & \multicolumn{2}{c}{\textit{room}} \\
                      & \textbf{GPU}       & \textbf{CPU}        & \textbf{GPU}         & \textbf{CPU}          & \textbf{GPU}         & \textbf{CPU}       \\
        \midrule
                1024$\times$1  &           23.14       &  786.53      &           20.99       &  518.14      &           20.38       &  527.26     \\
                 512$\times$2  &           12.80       &  500.43      &           11.57       &  334.03      &           11.32       &  321.02     \\
                 256$\times$4  &            7.45       &  326.37      &            6.73       &  222.21      &            6.68       &  211.15     \\
                 128$\times$8  &            4.98       &  221.90      &            4.47       &  157.40      &            4.51       &  149.24     \\
                 64$\times$16  &  \textbf{ 18.48}      &  175.64      &  \textbf{ 16.08}      &  130.48      &  \textbf{ 16.76}      &  122.29     \\
                 32$\times$32  &  \textbf{ 70.63}      &  150.91      &  \textbf{ 62.78}      &  115.97      &  \textbf{ 63.72}      &  106.26     \\
                 16$\times$64  &  \textbf{206.58}      &  141.55      &  \textbf{179.39}      &  111.31      &  \textbf{186.22}      &  103.05     \\

        \bottomrule
    \end{tabular}
    \label{tab:mstiming-windows}
\end{table}

\paragraph*{Mixing Gaussians with meshes in production rendering.}

In \cref{fig:seashell}, we show that our method enables the composition of Gaussians into a mesh-based environment, where the former project a soft shadow on the base, and can be seen from the refraction of the curved glass or the glossy reflection on the base or back panel. In addition, the path tracer can replicate the camera defocus blur accurately around the edges of the Gaussians.
In \cref{fig:drjohnsoncafe}, by putting a cafe brewing asset into the \emph{drjohnson} scene, we show that our method enables lighting a mesh-based asset with complex materials plausibly with an environment made of Gaussians. Both scenes are rendered with 1024 spp and then denoised with a production denoiser based on the work by I\c{s}\i k et al.~\cite{icsik2021interactive}, which works well since our method produces unbiased estimates. We use QMC sequences for camera-ray generation. Because our ray-tracing method allows combinations of assets, global stratified sampling across the entire scene becomes crucial: we observed that some of the noise originates from Gaussians, while other noise stems from solving the rendering equation on mesh surfaces.

\paragraph*{Limitations and future work.}

Our method can evaluate Gaussian splat opacity in various ways, and can approximately match a rasterizer by doing so using the projected camera depth. Minor differences to rasterization still remain in rendering results and could be reduced with further effort, but we believe that instead of carefully matching rasterization errors, it is better to invest in more accurate reconstruction.

Our current implementation assumes the radiance of Gaussians is known and unaffected by surrounding objects or lighting. However, our method could be combined with relightable Gaussians~\cite{r3dg,gsir} without changes to the core algorithm.

Finally, we believe our method can be extended to differentiable rendering; a straightforward gradient applied to the selected Gaussian from our method would be easy to compute, but lower-variance estimators could be derived with further research.

\section{Conclusion}
\label{sec:conclusions}

We presented a stochastic ray-tracing method to render large collections of transparent primitives such as 3D Gaussian. Instead of processing all ray-Gaussian intersections in sequential order, only a single BVH traversal finds all Gaussians potentially contributing to the ray. The opacity of each primitive is treated as a probabilistic decision between a fully opaque and fully transparent event, which means only the nearest opaque event needs to be found, avoiding the need for sorting. We show that the resulting Monte Carlo estimator is unbiased and the method has interactive performance even on low-end hardware.
Our method has been integrated in a commercial rendering product; we believe it can inspire further research at the intersection of Monte Carlo rendering, 3D capture and generation.

We believe our work can inspire further research at the intersection of Monte Carlo rendering, 3D capture, and generation.
Our method has been integrated into a feature-rich production renderer which is used in Adobe Substance 3D Viewer \cite{substance-viewer}.
Our experimental results were obtained using that renderer with comprehensive ray payloads, rather than a minimal, GS-specific implementation.


\bibliographystyle{eg-alpha-doi}
\bibliography{bibliography}


\end{document}